\documentclass[structabstract]{aa}  
\usepackage{lscape}
\usepackage[british]{babel}
\usepackage{natbib}
\bibpunct{(}{)}{;}{a}{}{,} 

\usepackage{graphicx}
\usepackage{url}
\usepackage{txfonts}
\begin{document}
\title{The Pisa pre-main sequence tracks and isochrones}

\subtitle{A database covering a wide range of Z, Y, mass, and age values \thanks{Tracks and isochrones are available on the web at the URL: http://astro.df.unipi.it/stellar-models/}\fnmsep\thanks{Tracks and isochrones are also available in electronic form at the CDS via anonymous ftp to cdsarc.u-strasbg.fr (130.79.128.5) or via http://cdsweb.u-strasbg.fr/cgi-bin/qcat?J/A+A/= } } 

\author{
  E. Tognelli \inst{1,2}
  \and
  P. G. Prada Moroni \inst{1,2}
  \and
  S. Degl'Innocenti \inst{1,2}
  }

\institute{
  Physics Department ``E. Fermi'', University of Pisa, Largo B. Pontecorvo 3, I-56127, Pisa, Italy\\ 
  \and
  INFN, Largo B. Pontecorvo 3, I-56127, Pisa, Italy\\
  }

\offprints{P. G. Prada Moroni, prada@df.unipi.it}
\date{Received 18 December 2009 / Accepted 4 July 2011}

\abstract
   {In recent years new observations of pre-main sequence stars (pre-MS) with $\mathrm{Z}\le \mathrm{Z}_\odot$ have been made available. To take full advantage of the continuously growing amount of data of pre-MS stars in different environments, we need to develop updated pre-MS models for a wide range of metallicity to assign reliable ages and masses to the observed stars.} 
   {We present updated evolutionary pre-MS models and isochrones for a fine grid of mass, age, metallicity, and helium values.}
   {We use a standard and well-tested stellar evolutionary code (i.e. \texttt{FRANEC}), that adopts outer boundary conditions from detailed and realistic atmosphere models.  In this code, we incorporate additional improvements to the physical inputs related to the equation of state and the low temperature radiative opacities essential to computing low-mass stellar models.}
   {We make available via internet a large database of pre-MS tracks and isochrones for a wide range of chemical compositions ($\mathrm{Z}=0.0002-0.03$), masses (M~=~$0.2-7.0$~M$_\odot$), and ages ($1-100$ Myr) for a solar-calibrated mixing length parameter $\alpha$ (i.e. 1.68). For each chemical composition, additional models were computed with two different mixing length values, namely $\alpha=1.2$ and 1.9. Moreover, for $\mathrm{Z}\ge 0.008$, we also provided models with two different initial deuterium abundances. The characteristics of the models have been discussed in detail and compared with other work in the literature. The main uncertainties affecting theoretical predictions have been critically discussed. Comparisons with selected data indicate that there is close agreement between theory and observation.}
   {}
   
\keywords{Stars: pre-main sequence -- Stars: evolution -- Stars: formation -- Stars: interiors  -- Stars: low-mass -- Stars: HR diagrams}

\maketitle

\section{Introduction}
We present a new set of pre-main sequence (pre-MS) stellar tracks and isochrones based on both the state-of-the-art input physics and outer boundary conditions. These models cover a large range of metallicities, helium abundances, masses, and ages, providing useful tools for interpreting observational data. Pre-MS tracks and isochrones collectively represent the theoretical tool needed to infer the star formation history and the initial mass function of young stellar systems. Thus, the availability of a large database with a fine grid of chemical compositions, masses, and ages will allow us to improve our understanding of the star formation process taking full advantage of the growing amount of data of very young clusters and associations in the Milky Way \citep[e.g.][]{stolte05,delgado07,brandner08}, the Small Magellanic Cloud \citep[e.g.][]{gouliermis06a,nota06,carlson07,gouliermis07b,sabbi07,gouliermis08,Cignoni09,cignoni10}, and the Large Magellanic Cloud \citep[e.g.][]{romaniello04,gouliermis06b,romaniello06,gouliermis07a,dario09}.

As an example of an intriguing application of pre-MS isochrones, \cite{Cignoni09} inferred the star formation history of NGC602, a very young stellar system in the Small Magellanic Cloud \citep[see also][]{nota06,carlson07,gouliermis07b,sabbi07}. 
 
From the theoretical point of view, as early understood by \citet{hayashi61} and \citet{hayashi63}, the evolution of hydrostatic pre-MS stars is in principle less complex than more advanced evolutionary phases from the numerical point of view, since no special algorithms are required to follow what it is essentially a quasi-static contraction. However, the pre-MS track location in the HR diagram strongly depends on the physical ingredients used in the evolutionary codes, such as the equation of state (EOS), the low-temperature radiative opacity, the outer boundary conditions, and the adopted convection treatment \citep[see e.g.,][]{dantona93,dantona94,DM97,BCAH98,SD00,baraffe02,montalban04b}. The uncertainties due to these quantities, progressively increase as the stellar mass decreases, especially for very low-mass stars (i.e. M~$\la 0.5$~M$_\odot$). Therefore, to compute models as accurately as possible, it is mandatory to
include the most recent updates of the above quoted physical inputs. In the past two decades, several generations of theoretical pre-MS models have been developed following improvements to the physical inputs to provide progressively more accurate theoretical predictions \citep[e.g.][]{BCAH98,DM97,SD00,dotter07,dicriscienzo09}.
 
The theoretical tracks and isochrones for pre-MS stars described in the present paper have been computed with an updated version of the \texttt{FRANEC} evolutionary
code \citep[][]{deglinnocenti08,valle09} relying on the most recent physical inputs. The models cover a wide range of metallicity ($\mathrm{Z}~=~0.0002-0.03$), masses (M $= 0.2 -7.0$M$_\odot$) and ages (t~=~$1-100$ Myr). The corresponding database is available on the web\footnote{Models and isochrones are available in the theoretical plane and will soon be available in different photometric systems.}.

The input physics adopted in the calculations are described in Sect. 2, followed by a discussion of the main sources of uncertainty in the evolution of young stars. In Sects. 3 and 4, our tracks and isochrones are compared with recent evolutionary models and data. Finally, in our last section the database is described.

\section{The models}
We describe the adopted physical inputs focusing only on the ingredients that affect mainly the pre-MS evolution and the location of the track in the HR diagram. 

Before starting this description, we recall that our models are standard in the sense that they do not include rotation, magnetic fields, and accretion. This is an important point to emphasize because the inclusion of these processes affects the evolution of pre-MS models introducing additional sources of uncertainty \citep[see e.g.,][]{pinsonneault90,siess97b,dantona00a,siess01,stahler04,baraffe09,hosokawa11}. 

Although inconsequential to pre-MS evolution, we also note that our current evolutionary tracks take into account microscopic diffusion. In particular, the diffusion velocities of He, Li, Be, B, C, N, O, and Fe are directly computed following the method described in \citet{thoul94}, while the other elements are assumed to diffuse as Fe but with a velocity scaled to their abundances. We do not consider radiative levitation.

\subsection{Equation of state}
\label{sec:eos}
To compute stellar models, it is necessary to use an accurate EOS covering the very wide range of temperatures and densities spanned during the entire evolution.
We adopted the most updated version of the OPAL EOS\footnote{The OPAL EOS tables are available at the URL: \url{http://opalopacity.llnl.gov/EOS_2005/}} in the version released in 2006 \citep{rogers96,rogers02}. 

\begin{figure}
\centering
\includegraphics[width=\linewidth]{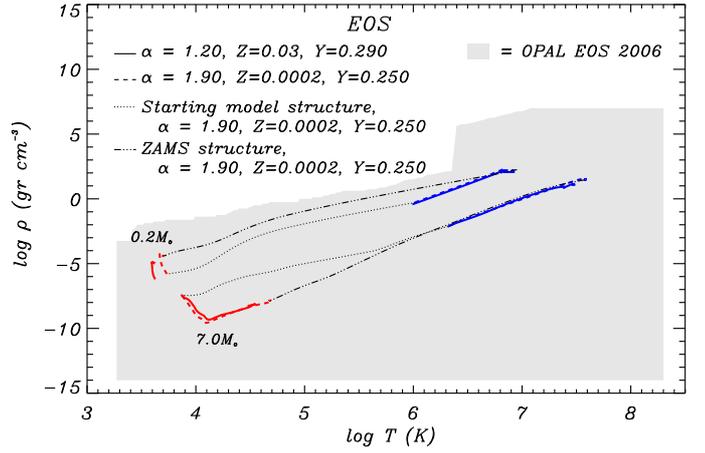}
\caption{Validity domain of the OPAL EOS (\textit{shaded area}) in the plane ($\log\,\mathrm{T}$, $\log\,\rho$); we superimposed the time evolution of the bottom of the atmosphere ($\tau_{\mathrm{ph}}$=10) and the stellar centre for $\mathrm{Z}=0.03$ (\textit{thick solid line}) and $\mathrm{Z}=0.0002$ (\textit{thick dashed line}), and the whole structure at the beginning of the contraction along the Hayashi track (\textit{dotted line}) and in the ZAMS (\textit{dot-dashed line}) for $\mathrm{Z}=0.0002$.}
\label{fig:eos_range}
\end{figure}

Figure \ref{fig:eos_range} shows the validity domain in the temperature-density plane of the OPAL EOS. Thick solid lines represent the time evolution during the pre-MS phase of the stellar centre (top-right side) and the atmosphere base at an optical depth\footnote{As we will describe in Sect. \ref{sec:BC} we choose $\tau_{\mathrm{ph}} =10$ for the matching point between the atmosphere model and the interior.} $\tau_{\mathrm{ph}}$=10 (bottom-left side) of the 0.2 M$_\odot$ and 7 M$_\odot$ models with $\mathrm{Z}=0.03$, $\mathrm{Y}=0.290$, and $\alpha=1.2$, which is our coldest set of tracks. To clearly show that the OPAL EOS is fully capable of covering the space of parameters required to compute all the models included in the database, we added to Fig. \ref{fig:eos_range} the time evolution of the central and outer regions of our hottest models, that is, those with $\mathrm{Z}=0.0002$, $\mathrm{Y}=0.250$, and $\alpha=1.9$ (thick dashed lines); in this case, the entire structure profiles for the starting model at the beginning of the pre-MS evolution (dotted lines) and for the final ZAMS model (dot-dashed lines) are also shown.  

\begin{figure}
\centering
\includegraphics[width=\linewidth]{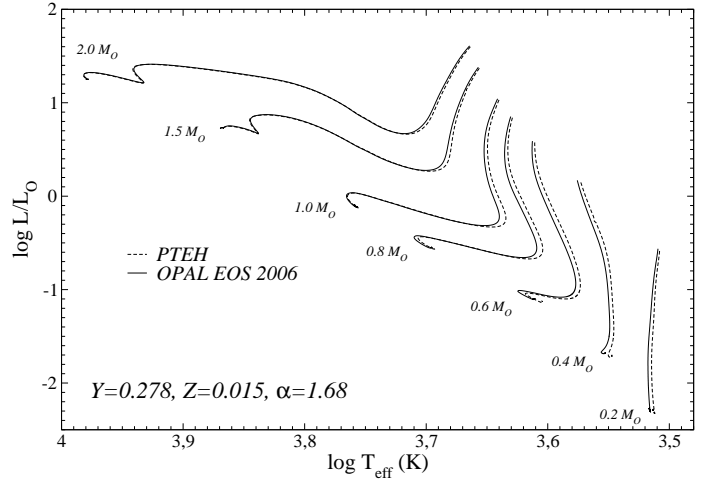}
\caption{HR diagram for evolutionary tracks in the mass range $0.2~-~2.0$~M$_\odot$ computed adopting the OPAL EOS 2006 (\textit{solid line}) and the FreeEOS in the PTEH configuration (\textit{dashed line}) for the labelled chemical composition and $\alpha$.}
\label{fig:eos_pteh} 
\end{figure}
Figure \ref{fig:eos_pteh} shows the effect of two different EOS widely used in the literature, namely the OPAL EOS 2006 and the FreeEOS \citep[for more details see][]{irwin04} in the configuration that should reproduce the PTEH EOS \citep{pols95}. The effect of the EOS has been widely discussed in the literature in the case of pre-MS objects \citep[see e.g,][ and references therein]{mazzitelli89,dantona93} and for low-mass main sequence stars \citep[see e.g.,][]{dorman89,neece84,chabrier97}; here we limit our discussion to presenting the results of our comparisons. For stars more massive than 1~M$_\odot$, the effect is essentially limited to the Hayashi track; the difference in effective temperature is about 50~K for 1~M$_\odot$ model and progressively decreases as the mass increases becoming about 30 K for 2~M$_\odot$ and almost negligible for more massive stars. For models less massive than 1~M$_\odot$, the EOS also affected the ZAMS because of thick convective envelopes present in the hydrogen burning phase, whose structure depends on the adiabatic gradient. For this mass range, the maximum differences between the tracks with OPAL and the PTEH EOS at fixed luminosity, shown in Table \ref{tab:diffe_eos}, are of about $30-40$~K along the Hayashi track and $30-60$~K in ZAMS depending on the mass. We note that for low-mass stars the ZAMS luminosity also depends on the adopted EOS.
 
\begin{table}
\caption{Differences between models computed adopting the PTEH and the OPAL EOS. The second column shows for each value of the mass the maximum difference at a fixed $\log \mathrm{L/L}_\odot$ along the Hayashi track. The third and forth columns show the differences in T$_{\mathrm{eff}}$ and $\log \mathrm{L/L}_\odot$, respectively, for ZAMS models.}
\label{tab:diffe_eos}      
\centering    
\begin{tabular}{c|c|c|c}
M/M$_\odot$ & \multicolumn{2}{|c|}{$\Delta \mathrm{T}_{\mathrm{eff}}$} & $\Delta \log \mathrm{L/L}_\odot$\\
\cline{2-4}
 & Hayashi & ZAMS &  ZAMS \\
 \hline
 0.2 & -30 K & -30 K & -0.02 \\ 
 0.4 & -35 K & -48 K & -0.04\\
 0.6 & -40 K & -60 K & -0.03\\
 0.8 & -46 K & -35 K & -0.01\\
 1.0 & -48 K & -15 K & - \\
 1.5 & -34 K & - & - \\
 2.0 & -30 K & - & - \\
\hline
\end{tabular}
\end{table}

\subsection{Boundary conditions}
\label{sec:BC}
To solve the differential equations of the stellar interior structure, suitable outer boundary conditions are required. The usual approach followed in standard evolutionary codes consists of taking as outer boundary conditions the values of the main physical quantities at the base of the atmosphere. These quantities are provided by a direct integration of the equations describing a mono-dimensional atmosphere both in hydrostatic equilibrium and for the diffusive approximation of radiative transport, in addition to a grey T($\tau$) relationship between the temperature and the optical depth $\tau$. The T($\tau$) relationships most commonly chosen are those of the Eddington approximation and \citet[][hereafter KS66]{krishna66}.

However, as discussed in several papers \citep[see e.g.,][]{auman69,dorman89,saumon94,baraffe95,allard97,chabrier97,BCAH98,baraffe02}, this technique is too crude for low-mass stars and in general for cold atmospheres. A much more sophisticated approach consists of adopting as outer boundary conditions for the integration of the stellar interior equations the main physical quantities at a given optical depth provided by detailed, non-grey atmospheric models that solve the full radiative transport equation.

We adopted the atmospheric models by \citet{brott}, hereafter BH05, computed with the PHOENIX code \citep{phoenix}, that are available for the parameter ranges $3000$ K $\le \mathrm{T}_{\mathrm{eff}}\le10\,000$ K, $-0.5\le~\log \mathrm{g}\,(\mathrm{cm\,s^{-2}}) \le +5.5$, and $-4.0 \le \mathrm{[M/H]} \le+0.5$, adopting $\alpha_{\mathrm{atm}}=2.0$ and the solar mixture of \citet{grevesse933}.

In the range $10\,000$ K $ \le \mathrm{T}_{\mathrm{eff}} \le 50\,000 $ K, $+0.0\le~\log \mathrm{g}\,(\mathrm{cm\,s^{-2}})\le+5.0$, and $-2.5 \le \mathrm{[M/H]} \le+0.5$, where the models of BH05 are unavailable, we used the atmospheric structures of \citet{castelli03}, hereafter CK03. These models are computed for the solar mixture of \citet{grevesse98}. As in the previous case, the mixing-length scheme is followed to describe the convection, but with a lower value of the mixing-length parameter $\alpha_{\mathrm{atm}}=1.25$. The effect of the adoption of two different $\alpha_{\mathrm{atm}}$ values should be small compared to the other uncertainty sources, since for effective temperatures above $10\,000$ K, which roughly correspond to ZAMS stars of M $\ga$ 2.0 M$_\odot$, the convective envelope, when present, is very thin. 

From these two sets of atmospheric models, we built tables of boundary conditions for a fixed value of $\tau_{\mathrm{ph}}$, defined as the optical depth at which the interior and atmosphere matches. In more detail, these tables contain the temperature T($\tau_{\mathrm{ph}}$, T$_{\mathrm{eff}}$, $\log \mathrm{g}$, Z) and the pressure P($\tau_{\mathrm{ph}}$, $\mathrm{T}_{\mathrm{eff}}$, $\log \mathrm{g}$, Z) at the optical depth $\tau_{\mathrm{ph}}$ extracted from the atmospheric models for the available T$_{\mathrm{eff}}$, $\log \mathrm{g}$, and Z. During the computation of the stellar structure and evolution, a spline interpolation of these tabulated values at the $\mathrm{T}_{\mathrm{eff}}$, $\log \mathrm{g}$ and $\log \mathrm{Z}$ of the star is performed to obtain the corresponding T($\tau_{\mathrm{ph}}$) and P($\tau_{\mathrm{ph}}$) required for the interior integration. For our reference set of models we chose the commonly suggested value of $\tau_{\mathrm{ph}} = 10$ \citep{morel94}; this value represents a good compromise between two opposite requirements, namely, a large $\tau_{\mathrm{ph}}$ to guarantee the validity of the photon diffusion approximation, and a low $\tau_{\mathrm{ph}}$ to avoid large discrepancies introduced by the inconsistencies often present between atmospheric and interior models. We note that we performed the interpolation in the global metallicity Z without taking into account the different heavy element distributions in the interior and atmospheric regions.

It is worth noting that the interpolation technique adopted to obtain the boundary conditions must be chosen carefully, because the available atmospheric models are not usually computed with this in mind and the resolution in $\mathrm{T}_{\mathrm{eff}}$, $\log \mathrm{g}$, and [M/H] is often not very high. This might cause problems, mainly in the region of the HR diagram populated by red giant stars, in which the thickness of the convective outer layer changes abruptly varying $\mathrm{T}_{\mathrm{eff}}$ and $\log \mathrm{g}$. As an example, we verified that if a spline interpolation is replaced with a linear one, large and non-physical waving appears in the HR diagram along the Hayashi phase.

\subsection{Opacity}
\label{sec:opacity}
We adopted the version of the OPAL\footnote{The OPAL radiative opacity can be found at the URL: \url{http://opalopacity.llnl.gov/new.html}} radiative opacity tables \citep[see e.g. ][]{iglesias96} released in 2005 for $\log\,\mathrm{T(K)}>4.5$; for this range of temperatures, the radiative opacity and the EOS are fully consistent, since they have been computed by the same group adopting the same physical prescriptions. For lower temperatures, characteristic of the outer stellar regions, we used the radiative opacity of \citet{ferguson05} (hereafter F05)\footnote{The F05 low temperature radiative opacities are available at the URL: \url{http://webs.wichita.edu/physics/opacity}}, which include an accurate description of the molecular absorptions; indeed for $\mathrm{T}\le4500$ K, hence for low-mass stars, molecules become the main source of opacity. We also mention, for the sake of completeness, that we used the conductive opacity of \citet{potekhin99} and \citet{shternin06} (hereafter PSY06)\footnote{PSY06 conductive opacities are available at the URL: \url{http://www.ioffe.ru/astro/conduct/} }; however, for the present calculations thermal conduction is completely negligible, and we verified that even our lowest mass models are unaffected by increasing the conductive opacity of $5\%$. 

Our reference set of models was computed adopting the solar heavy element distribution of \citet{asplund05} in both the low and high temperature regimes.
           
\begin{figure}
\centering
\includegraphics[width=\linewidth]{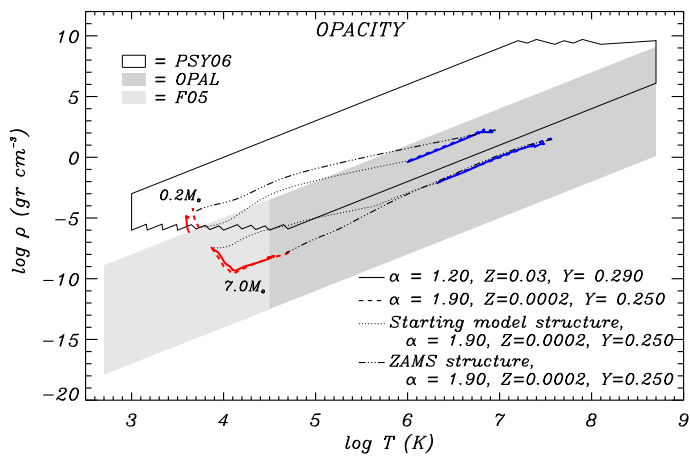}
\caption{Validity domain of the radiative opacity tables (OPAL and F05, \textit{shaded area}) and of the conductive ones (\textit{solid box}). As in Fig. \ref{fig:eos_range} we superimposed the time evolution of the bottom of the atmosphere ($\tau_{\mathrm{ph}}$=10) and the stellar centre for $\mathrm{Z}=0.03$ (\textit{thick solid line}) and $\mathrm{Z}~=~0.0002$ (\textit{thick dashed line}), and the whole structure for the first model on the Hayashi track (\textit{dotted line}) and for the ZAMS (\textit{dot-dashed line}) for $\mathrm{Z}=0.0002$.}
\label{fig:kappa_range}
\end{figure}
Figure \ref{fig:kappa_range} shows the range of validity in the temperature-density plane of the F05, OPAL, and PSY06 opacity tables. we note that the actual validity domain of the OPAL tables extends down to $\log \mathrm{T(K)} = 3.75$, partially overlapping with the F05 tables, but we do not show it in figure, since we used the F05 opacities for $\log \mathrm{T(K)} \le 4.5$. As in Fig. \ref{fig:eos_range}, we plotted the time evolution  during the pre-MS phase of the stellar centre (top-right side) and the atmosphere base at an optical depth $\tau_{\mathrm{ph}}$=10 (bottom-left side) of the 0.2 M$_\odot$ and 7 M$_\odot$ models with $\mathrm{Z}=0.03$, $\mathrm{Y}=0.290$, and $\alpha=1.2$, i.e. our coldest set (thick solid line). The figure also shows the time evolution of the central and outer regions of our hottest models, that is, those with $\mathrm{Z}=0.0002$, $\mathrm{Y}=0.250$, and $\alpha=1.9$ (thick dashed lines); in this case, the entire structure profiles for the starting model at the beginning of the pre-MS evolution (dotted lines) and for the final ZAMS model (dot-dashed lines) are also shown.
 
We note that the current version of the radiative opacity tables do not extend sufficiently into the temperature-density plane to cover the entire pre-MS evolution of stars less massive than $0.6- 0.5$ M$_\odot$ (the exact value depending on the chemical composition), hence extrapolation to higher densities is required for both OPAL and F05 opacity tables. To our present knowledge, there are no opacity tables available in the literature that cover the whole range of temperatures and densities spanned by low-mass pre-MS models. Moreover, the extrapolation of radiative opacities is a dangerous procedure, because opacity coefficients are very sensitive to temperature and density. Given this situation, it is of primary importance to evaluate the sensitivity of the computed evolution to the adopted extrapolation method.

We tested three different methods: constant, linear for the last two points, and a linear fit to the last four points extrapolation. The reason for trying this latter method is to extend the mean slope of the opacity with respect to the density to reduce the contribution of a possible jump in the opacity near the border.
 
We performed test of both non-grey and grey models, the latter ones having been computed adopting the classical T($\tau$) relationship of \citet{krishna66}. We found that the extrapolation method does not affect the non-grey models, because the extrapolated values of the radiative opacity are used only in a very thin shell, but not for computing the boundary conditions, which are provided by atmospheric models. In contrast, low-mass grey models actually use the extrapolated radiative opacity coefficients over the whole outermost region, atmosphere included, making the structures much more sensitive to the adopted extrapolation method.

\begin{figure}
\centering
\includegraphics[width=\linewidth]{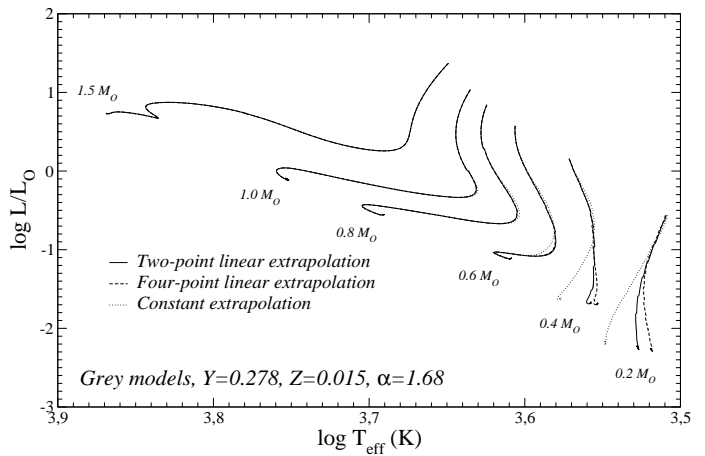}
\caption{Comparison in the HR diagram of low-mass tracks computed adopting three different methods of opacity extrapolation (see text) in the regions not currently  covered by opacity tables (low temperature-high density, see Fig. \ref{fig:kappa_range}). }
\label{fig:extra}
\end{figure}

Figure \ref{fig:extra} shows the HR diagram of grey pre-MS models computed with the three different extrapolation techniques. The impact of the extrapolation is completely negligible for stars more massive than $0.6$~M$_\odot$, because their structures are entirely covered by the OPAL+F05 tables, and becomes progressively larger and larger as the mass decreases, becoming quite important for stars less massive than  $0.4$ M$_\odot$. The constant extrapolation, which is the most crude possibility, leads to  ZAMS models hotter than those computed with a linear fit extrapolation of about 170~K  and 160~K for 0.4 M$_\odot$ and $0.2$ M$_\odot$, respectively. The difference between linear extrapolation with two or four points is lower (by about $45-60$~K).

\subsection{Solar mixture of heavy elements}
\label{sec:mixture}
A quite important, although often not explicitly specified, ingredient of stellar modelling is the heavy element mixture, which is the number distribution of chemical elements heavier than helium. The adopted mixture affects both the radiative opacity of the stellar matter and the nuclear burning efficiency, through, for example, the C, N, O abundances \citep[see e.g. ][]{scilla06,sestito06}. The heavy element mixture usually adopted is that of the Sun for population I stars, while for population II stars a given amount of $\alpha$-element enhancement (for the same mixture) is taken into account. The precise values of the solar chemical abundances have become controversial because of the considerable impact of the new 3D hydrodynamic atmospheric models \citep{asplund05,asplund09,caffau10}. 

We computed the stellar tracks currently available in the database with the mixture of \citet{asplund05} (hereafter AS05) before the revised version of the same group \citep[][hereafter AS09]{asplund09} was released; thus we assumed a reference value of solar photospheric metallicity of $\mathrm{(Z/X)}_{\mathrm{ph}}~=~0.0165$. Since this is one of the main novelty of the present calculations with respect to other classical pre-MS models, we analysed the effect of the adopted solar mixture by comparing models computed adopting three different distributions, namely, the widely used \citet{grevesse933} (hereafter GN93) and the two more recent AS05 and AS09. In each case, the same distribution was coherently used in both the opacity tables (both the OPAL and the F05) and the computation of the nuclear-burning energy release.

\begin{figure}
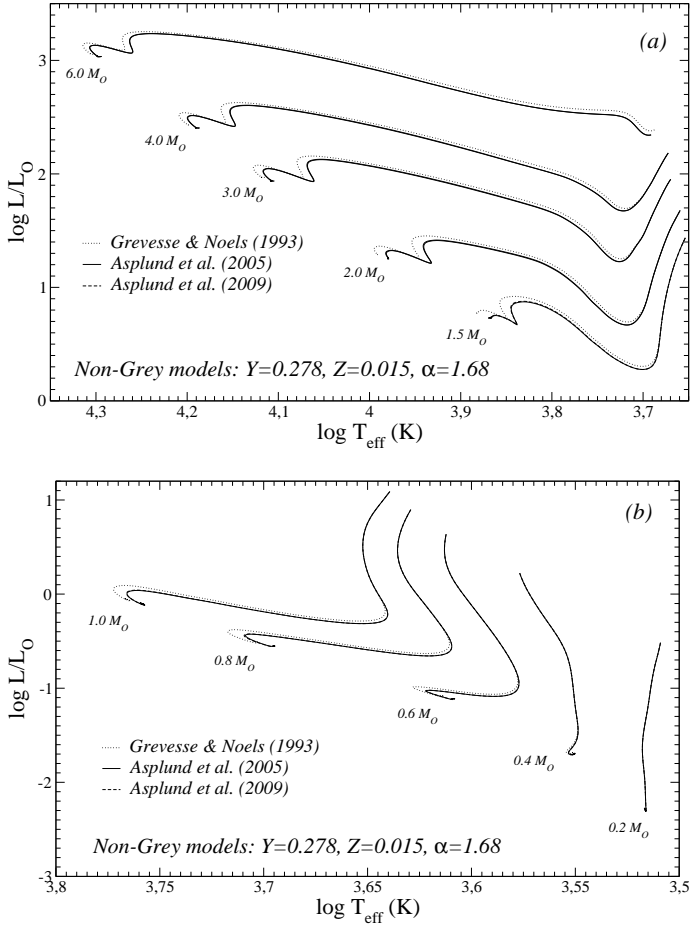

\centering
\includegraphics[width=\linewidth]{MIXTURE_BC_1.eps}\hspace{0.1cm}

\centering
\includegraphics[width=\linewidth]{MIXTURE_BC_2.eps}
\caption{Comparisons between non-grey models computed adopting the GN93, AS05 and AS09 heavy elements mixtures. The upper panel shows the mass range $1.5-6.0$~M$_\odot$, while the bottom one the range $0.2-1.0$~M$_\odot$.}
\label{fig:mistura_nongrey}
\end{figure}

Figure \ref{fig:mistura_nongrey} shows non-grey tracks computed with the quoted mixtures. Unfortunately, the atmospheric models adopted to obtain the boundary conditions are currently available for only one solar mixture (GS98 in CK03 and GN93 in BH05), hence the mixture is kept fixed in the atmosphere.

As clearly shown in Fig. \ref{fig:mistura_nongrey}, the models computed with the two releases of the solar mixture of Asplund are almost indistinguishable. This is unsurprising because AS05 and AS09 are quite similar, with in particular almost identical global abundances in mass of both the iron group and $\alpha$ elements at a given metallicity Z. 

In contrast, the tracks computed with the GN93 mixture become dissimilar to the others when a sizeable radiative core develops and the stars leave the Hayashi line moving towards higher effective temperatures. In the ZAMS, the GN93 models are hotter and brighter than the AS05 ones, any differences becoming smaller and smaller as the mass  decreases: $\Delta \mathrm{T}_{\mathrm{eff}} \approx$~300~K and $\Delta \log \mathrm{L/L}_\odot\approx 0.02$ for 6.0~M$_\odot$, $\Delta \mathrm{T}_{\mathrm{eff}}\approx $ 90~K and $\Delta \log \mathrm{L/L}_\odot\approx 0.05$ for 1 M$_\odot$, and negligible for 0.2~M$_\odot$. 

We note that the almost coincidence of the GN93 models with the others along the Hayashi track is the consequence of adopting the same heavy element mixture in the outermost layers, whose radiative opacity governs the effective temperature of the Hayashi track, because the outer boundary conditions were computed with the same atmosphere models (see discussion in Sect. \ref{sec:BC}).

Thus, to actually use different mixtures also in the atmosphere, we computed three additional sets of grey models by adopting the KS66 T($\tau$) relationship with $\tau_{\mathrm{ph}}=2/3$, even though grey boundary conditions should not be used in very low-mass stellar models (see e.g. Sect. \ref{sec:BC} and references therein). 

\begin{figure}[t]
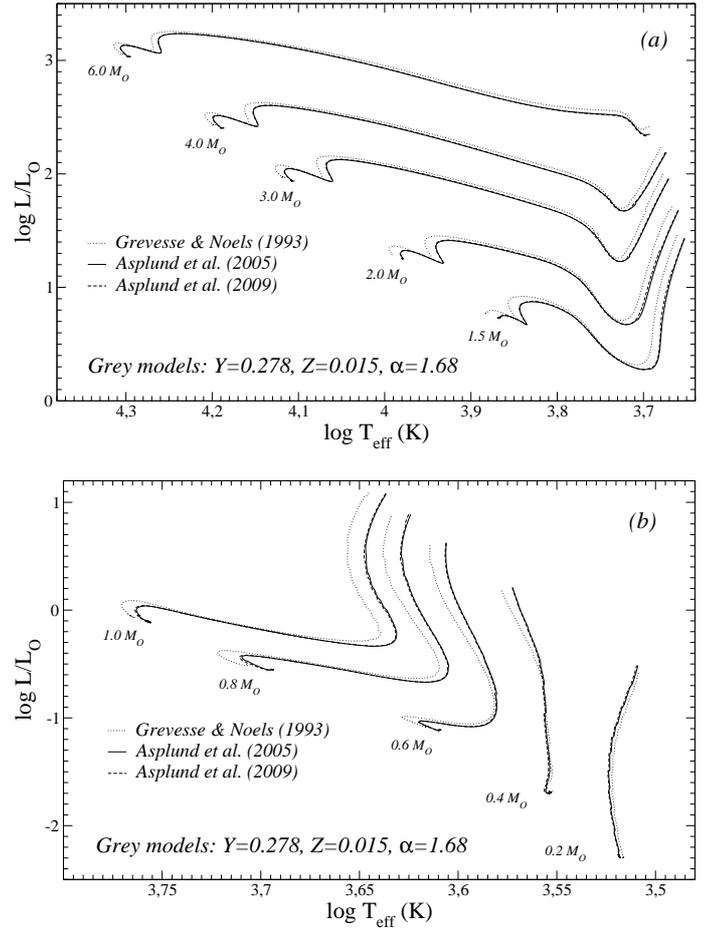

\centering
\includegraphics[width=\linewidth]{MIXTURE_GREY_1.eps}\hspace{0.1cm}

\centering
\includegraphics[width=\linewidth]{MIXTURE_GREY_2.eps}
\caption{As in Fig. \ref{fig:mistura_nongrey} but for grey models.}
\label{fig:mistura_grey}
\end{figure}

Figure \ref{fig:mistura_grey} shows the same kind of comparison as Fig. \ref{fig:mistura_nongrey} but for the grey models. As one can see, the GN93 models in this case  differ from the others also along the Hayashi track, as a consequence of having a different radiative opacity also in the outermost stellar layers that affects the outer boundary condition.

For masses greater than 0.4~M$_\odot$ the GN93 models are the hottest along the whole pre-MS evolution. For these masses, an increase in the iron abundance causes a larger radiative opacity, thus a lower effective temperature and luminosity, exactly what occurs passing from GN93 to AS05 for the same metallicity Z. For less massive stars,  the effect of updating the heavy element mixture from GN93 to AS05 on the pre-MS track is the result of two opposite trends: an increase in the H$^{-}$ contribution to the radiative opacity, as a consequence of the higher abundance of electron donors, and a decrease in the molecular opacity caused by the reduced oxygen abundance.

The previous discussion refers to solar metallicity models, although we also performed comparisons for a lower metallicity, i.e. $\mathrm{Z}=0.0002$, for the three solar mixtures discussed above in the case of non-grey models. For metal-poor stars, we found that the pre-MS tracks are insensitive to the heavy-element mixture in the HR diagram.

\subsection{Convection} 
The boundaries of the convectively unstable regions are fixed by the classic Schwartzschild criterion; within these boundaries, we assume that the mixing timescale is short compared to the nuclear burning one. 

As is well known, one major shortcoming of stellar modelling is the oversimplified treatment of convection, particularly in giant structures, such as RG and pre-MS stars, where large superadiabatic regions are present. The practical consequence is that we are not yet able to firmly predict the $\mathrm{T}_{\mathrm{eff}}$ and the radius of stars with an outer convective envelope, such as pre-MS objects during the Hayashi track. The common approach to stellar computation is to adopt the mixing length theory \citep{bohm58} for which the average convective efficiency depends on the mixing length $\ell=\alpha\,\textrm{H}_{\mathrm{p}}$, where $\mathrm{H}_{\mathrm{p}}$ is the pressure scale height and $\alpha$ is a free parameter that is calibrated with observations. We implemented the mixing length scheme in the code following the prescriptions of \citet{cox}.

The usual procedure for calibrating the mixing-length efficiency consists of constraining the $\alpha$ value by fitting the solar radius. For this reason, it might be useful to provide a brief description of the main characteristics that a standard solar model (SSM) must satisfy. A SSM is a stellar model of 1 M$_\odot$ such that at the age of the Sun (i.e. $\approx4.57$~Gyr) reproduces the observed solar luminosity, radius, and photospheric $\mathrm{(Z/X)}_{\mathrm{ph}}$. For this last quantity, we adopted the value 0.0165 from the heavy element mixture of \citet{asplund05}, the same mixture used in our reference models. In our SSM, the accuracy of the fit is very high, namely $\Delta \mathrm{L/L}<10^{-5}$, $\Delta \mathrm{R/R} <10^{-4}$, and $\Delta \mathrm{(Z/X)}_{\mathrm{ph}}/\mathrm{(Z/X)}_{\mathrm{ph}} < 4\cdot10^{-4}$. The parameters to tune to achieve such a good fit are the initial metallicity Z, helium abundance Y, and the mixing-length parameter $\alpha$.  We note that the observed value of the photospheric $\mathrm{(Z/X)}_{\mathrm{ph}}$ does not represent the original one, since microscopic diffusion reduces the helium and heavy-element surface abundances,  increasing the hydrogen one. 
 
\begin{figure}
\centering
\includegraphics[width=\linewidth]{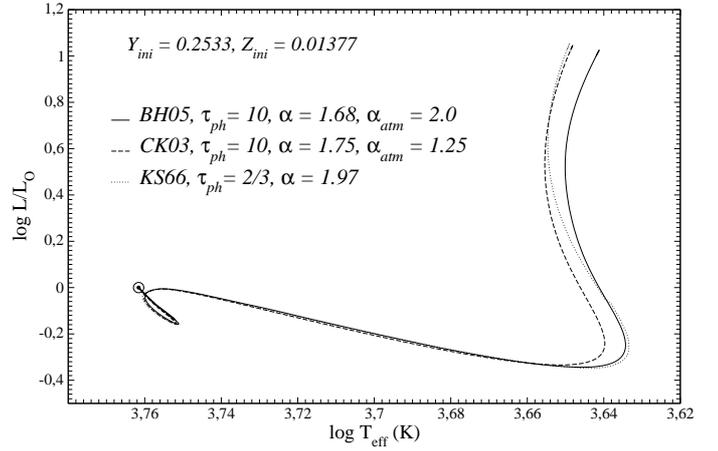}
\caption{Comparisons between SSM evolutionary tracks computed adopting the different labelled boundary conditions, namely a grey atmosphere plus the $\mathrm{T}(\tau)$ relationship by \citet{krishna66} (KS66, \emph{dotted line}) and the non-grey atmospheres by \citet{brott} (BH05, \emph{solid line}) and \citet{castelli03} (CK03, \emph{dashed line}) connected to the interior at $\tau_{\mathrm{ph}}=~10$. The symbol $\odot$ marks the position of the Sun.}
\label{fig:Sole}
\end{figure}
Figure \ref{fig:Sole} shows the evolutionary tracks corresponding to three different SSMs provided by the current version of the \texttt{FRANEC} code ($\mathrm{Y}_\odot=0.2533$, $\mathrm{Z}_\odot=0.01377$). The physical ingredients adopted to compute these models are exactly the same with the exception of the outer boundary conditions. In one case (dotted line), a simple grey atmosphere with the KS66 T($\tau$) relationship is used, while in the other two cases a non-grey detailed atmosphere models by \citet{brott} (solid line) and \citet{castelli03} (dashed line) are adopted.  

To fit the present Sun, the three models require three different values of the MLT parameter, namely $\alpha=1.97$ with the KS66 grey T($\tau$) relation, $\alpha=1.68$ and $\alpha=1.75$ with, respectively, BH05 and CK03 atmospheric models\footnote{Note that the mixing length parameter in the atmosphere $\alpha_{\mathrm{atm}}$ differs from that used in the interior, namely $\alpha_{\mathrm{atm}} = 2.0$ and $\alpha_{\mathrm{atm}} = 1.25$ for BH05 and CK03, respectively.}. 
As a result, the three models have a different pre-MS location and shape in the HR diagram, mainly near the Hayashi track, where the CK03 model is hotter than the BH05 one by about 60 K. The grey model is almost coincident with the CK03 one during the first contraction, while it moves towards the BH05 track as the star evolves. Similar results have already been discussed in the literature \cite[see e.g.,][]{dantona94,montalban04b}.
     
Although a better approach is still not possible, we recall that one should be careful when extending the solar-calibrated $\alpha$ to other stars because, as discussed in several papers, this could introduce systematic errors \citep[see e.g.,][]{pedersen90,canuto92,dantona94,salaris96,montalban04b}. We also note that there is no reason throughout the whole evolution of the star to assume a constant value of $\alpha$ \citep[see e.g.,][and references therein]{siess97,piau11}. In this respect, the progresses in 2D and 3D radiative hydrodynamical simulations of convection in the past decade \citep{ludwig99,trampedach99,trampedach07} provided evidence of both variations in  $\alpha$, which can depends on both $\mathrm{T}_{\mathrm{eff}}$ and $\log \,\mathrm{g}$, and the possibility of theoretically calibrating the mixing length parameter \citep[see also][]{montalban06}. Moreover, there appears to be a lower efficiency of the superadiabatic convection in low-mass stars ($\alpha \approx 1$) with respect to the solar-calibrated one according to observations of some binary systems \citep[see e.g.,][]{simon00,steffen01,stassun04}, and $^7$Li data of young open clusters \citep[][]{ventura98,dantona03}. 

This uncertainty in the mixing length efficiency also affects both mass and age estimates of young stars, since these are often obtained by comparing the stellar position in the HR diagram with the theoretical tracks, whose effective temperature depends on $\alpha$ whenever a convective envelope is present.
    
\begin{figure}
\centering
\includegraphics[width=\linewidth]{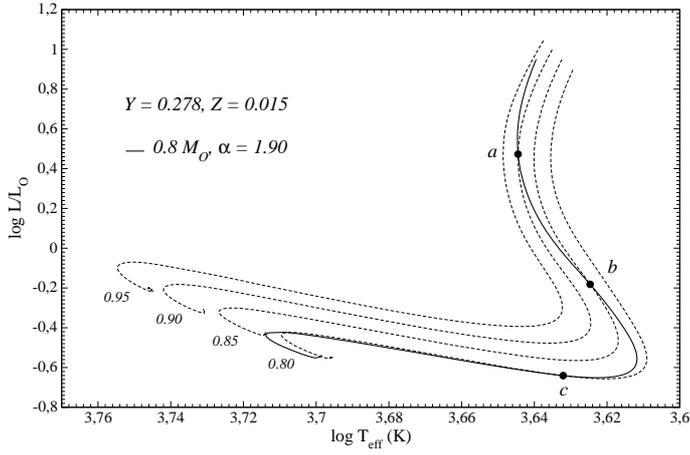}
\caption{Figure shows that the same position in the HR diagram can be reached by stars of different masses if the $\alpha$ parameter is changed.  This effect is shown with respect to a 0.8~M$_\odot$ star model with $\alpha=1.9$ (\textit{solid line}) by comparison with models with $\alpha=1.68$ and masses in the range $0.8-0.95$ M$_\odot$ (\textit{dashed lines}).  The boundary conditions are the same as our reference set of models.}
\label{fig:conv}
\end{figure}

Figure \ref{fig:conv} shows the pre-MS tracks of a 0.8 M$_\odot$ model with $\alpha=1.90$ and of four other masses in the range $0.8- 0.95$ M$_\odot$ with the solar-calibrated mixing length parameter, i. e. $\alpha=1.68$. The points mark the intersection between the first tracks and the other four. As expected, the greatest differences occur near the Hayashi track at the point '\textit{a}' where the same luminosity and temperature of the 0.8 M$_\odot$ $\alpha=1.9$ model is reproduced by a 0.9~M$_\odot$ models with $\alpha=1.68$. The different choice of $\alpha$ translates into differences in effective temperature of $90-100$ K, which correspond to uncertainties of about $0.1-0.15$ M$_\odot$ in the first phase of pre-MS contraction. This is consequently reflected in an uncertainty in age, because stars of different masses evolve on different timescales; the largest effect on age occurring at the point '\textit{b}'  where the relative difference is about $12\%$, whereas at the points '\textit{a}', '\textit{c}' is of the order of $6-7\%$.
 
Given this intrinsic weakness in the mixing-length solar calibration and the great impact of the efficiency of super-adiabatic convection, we decided to provide pre-MS models with three values of $\alpha$, namely 1.2, 1.68 (standard solar model obtained using the AS05 mixture and BH05 boundary conditions), and 1.9. 

\subsection{Nuclear network and deuterium burning}
\label{sec:deuterium}
The current version of \texttt{FRANEC} code follows the burning of 26 elements, in particular the burning of light elements (D,$^3$He, $^6$Li, $^7$Li, $^9$Be, and $^{11}$B) starting from the pre-MS phase, with initial abundances of $^3$He, $^6$Li, $^7$Li, $^9$Be, and $^{11}$B from \citet{geiss98}. The cross-sections of the nuclear reactions relevant to the hydrogen ($pp$ chain and $CNO$ cycle) and light element burning are assumed to be those of the NACRE compilation \citep{nacre}, with the exception of that of $^{14}$N(p,$\gamma$)$^{15}$O, which is from the LUNA collaboration \citep{imbriani05,luna06,lemut06}.

While the burning of Li, Be, and B does not contribute much to the stellar luminosity because of their low abundances, deuterium plays a crucial role during the first evolution of the pre-MS stars. The energy released by D-burning temporarily decelerates the star's contraction, which accelerates again once this element has been exhausted. The observable effect of D-burning is to increase the stellar luminosity for a given age in the first few millions years ($1-10$ Myr, depending on the stellar mass).

Recent estimates based on the WMAP results \citep[see e.g.][]{bennett03} and the standard model of cosmological nucleosynthesis, found a numerical abundance of the primordial deuterium relative to hydrogen (D/H)$_{\mathrm{p}}$~=~$2.75^{+0.24}_{-0.19}\cdot 10^{-5}$ \citep[see e.g. ][]{cyburt04}\footnote{By (D/H) we mean the ratio of the numerical abundances of deuterium to hydrogen.} that corresponds to a fractional abundance in mass in the range $3.8\cdot 10^{-5}\la \mathrm{X}_{\mathrm{D}} \la~4.5\cdot 10^{-5}$. A similar D-abundance was found by \citet{pettini08} and \citet{steigman07}. On the other hand, for the local ISM the value is lower, as a consequence of stellar astration. \citet{geiss98} found (D/H)$_{\mathrm{ISM}}~\approx2.1\cdot 10^{-5}$ ($\mathrm{X}_{\mathrm{D}}\approx 3\cdot 10^{-5}$), while more recent observations give (D/H)$_{\mathrm{ISM}} =1.88\,\pm\,0.11\cdot 10^{-5}$ that is $\mathrm{X}_{\mathrm{D}}\approx 2.6\cdot 10^{-5}$ \citep[see also][]{vidal98,linsky98,linsky06, steigman07}. 

Thus, $\mathrm{X}_{\mathrm{D}}=4\cdot10^{-5}$ and $\mathrm{X}_{\mathrm{D}}=2\cdot10^{-5}$ should be suitable for, respectively, population II and population I pre-MS stellar models.

\begin{figure}[t]
\centering
\includegraphics[width=\linewidth]{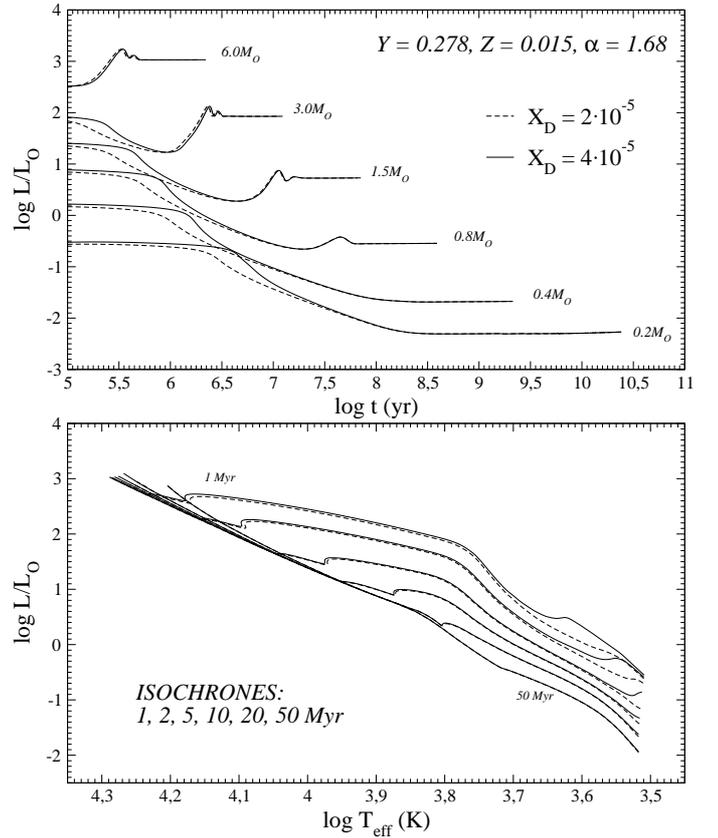}
\caption{Effects of two different assumptions about the initial deuterium abundance, namely $\mathrm{X}_{\mathrm{D}}=2\cdot 10^{-5}$ (\textit{dashed line}) and $\mathrm{X}_{\mathrm{D}}=4\cdot 10^{-5}$ (\textit{solid line}). Upper panel: time evolution of the luminosity for tracks in the mass range 0.2$-$6 M$_\odot$. Lower panel: HR diagram for isochrones of ages 1, 2, 5, 10, 20, and 50 Myr.}
 \label{fig:D_L_AGE_ISO}
\end{figure}
Figure \ref{fig:D_L_AGE_ISO} shows pre-MS models (upper panel) and isochrones (bottom panel) computed with $\mathrm{X}_{\mathrm{D}}=4\cdot 10^{-5}$ and  $\mathrm{X}_{\mathrm{D}}=2\cdot 10^{-5}$. As expected, the duration of the deceleration caused by the D-burning gets shorter with decreasing initial deuterium abundance. The extent of this effect depends on the mass of the star, since the deuterium exhaustion occurs at progressively earlier stages as the stellar mass increases. In particular, for stars with M~$\ga$~3.0~M$_\odot$, the differences in the evolutionary times are limited to very young ages, well below 1~Myr. Thus, in this mass range, the two sets of isochrones converge for ages older than 1~Myr, in particular for a 0.2~M$_\odot$ model, the effect of the adopted initial deuterium abundance is not negligible until much older ages, namely 10~Myr. We also note that we discuss is the amount of deuterium actually left by the previous protostellar evolution in Sect. \ref{sec:initial}.

\begin{figure}
\centering
\includegraphics[width=\linewidth]{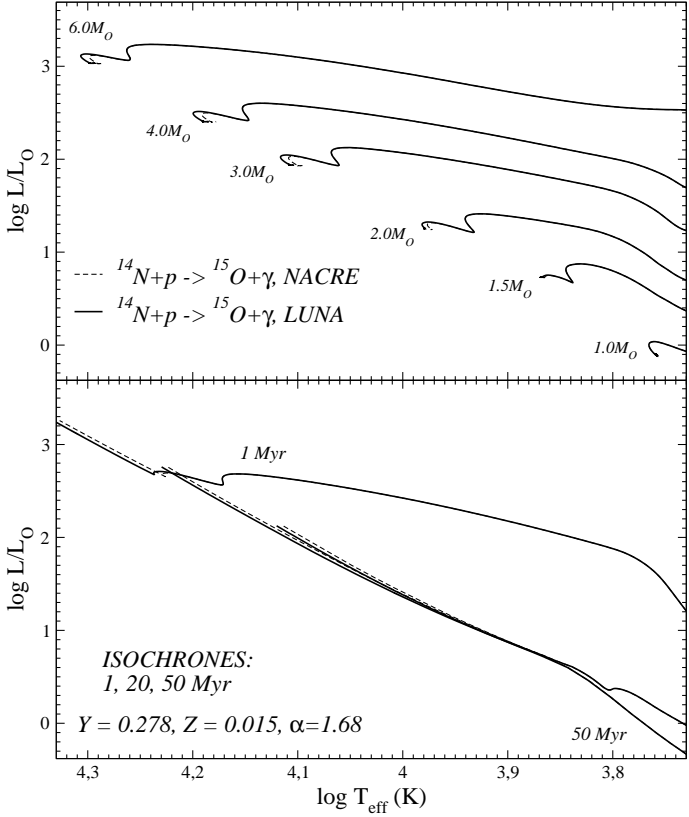}
\caption{Comparison between models with $\mathrm{Z}=0.015$, computed adopting the $^{14}$N(p,$\gamma$)$^{15}$O reaction rate from LUNA (\textit{solid line}) and NACRE (\textit{dashed line}). Upper panel: evolutionary tracks in the HR diagram in the mass range 1$-$6 M$_\odot$. Bottom panel: isochrones for 1, 20 and 50 Myr.}
\label{fig:azoto}
\end{figure}
Since one of the novelties of the current set of models is the $^{14}$N(p,$\gamma$)$^{15}$O cross-section provided by the LUNA collaboration, a detailed discussion of the changes caused by replacing the older but still widely used NACRE cross-section with the new one might be of some interest. The $^{14}$N(p,$\gamma$)$^{15}$O is the slowest reaction of the $CNO$ cycle and thus governs the efficiency of the central H-burning in stars more massive than $1.2-1.5$ M$_\odot$ (the exact value depending on the chemical composition), in which the \textit{CNO} cycle is dominant with respect to the $pp$ chain. 
 
Figure \ref{fig:azoto} shows a set of models and isochrones with $\mathrm{Z}=~0.015$ computed adopting the NACRE (dashed line) and the LUNA (solid line) $^{14}$N(p,$\gamma$)$^{15}$O cross-sections. As expected, the differences between the two sets of models become not negligible only for stars more massive than $1.2-1.5$ M$_\odot$ when they approach the ZAMS, that is, when the \textit{CNO} cycle becomes a significant source of energy. The ZAMS models computed with the LUNA cross-section are hotter than those computed with the NACRE one, the former being smaller than the latter. The greatest differences in $\mathrm{T}_{\mathrm{eff}}$ occur for higher masses ($\Delta \mathrm{T}_{\mathrm{eff}}\approx$~350~K for 6~M$_\odot$), while they decrease to 250~K for 3~M$_\odot$ and to $40$ K for 1.5~M$_\odot$.

\begin{figure}
\centering
\includegraphics[width=\linewidth]{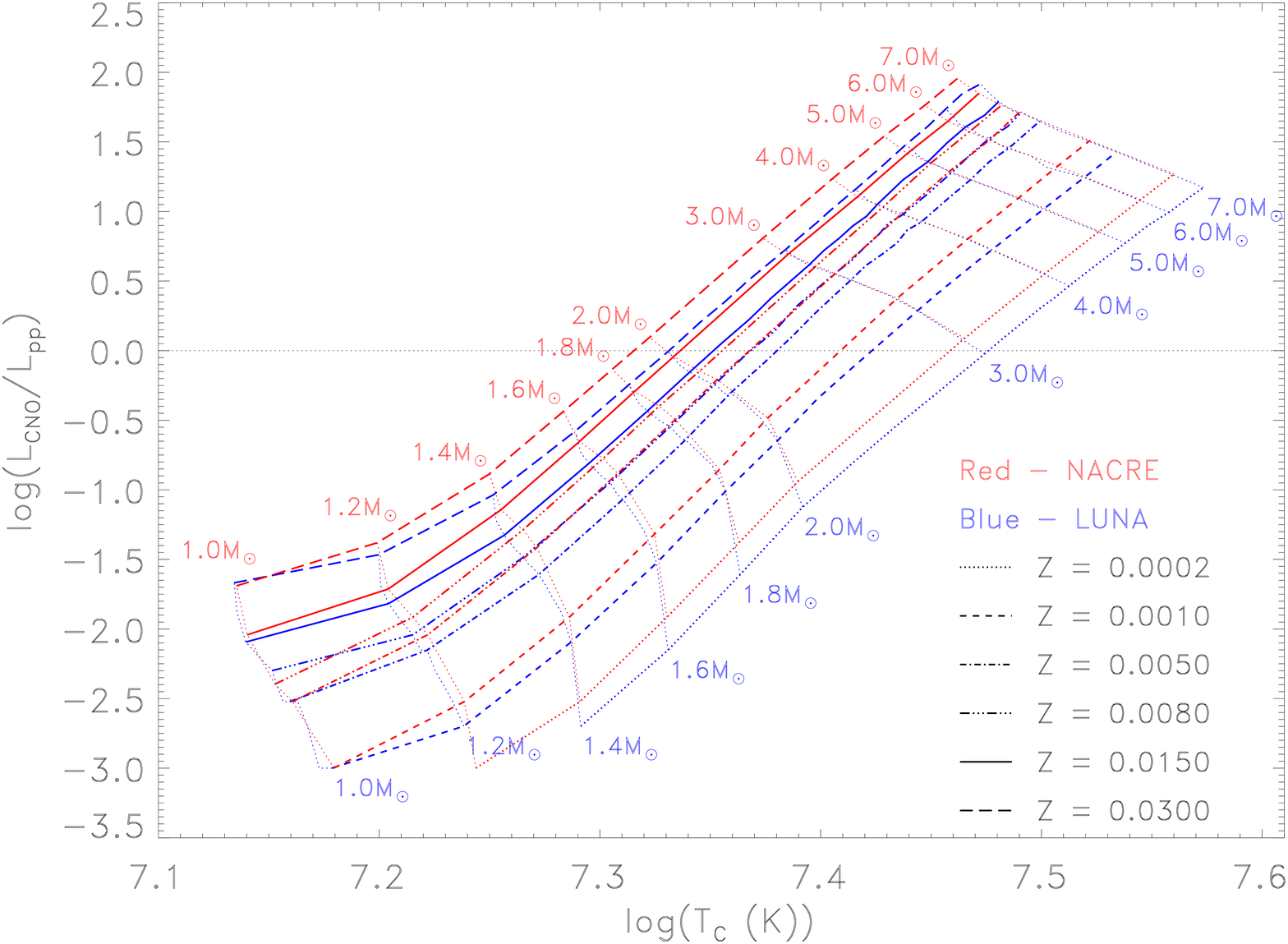}
\caption{Ratio of the $CNO$ to $pp$ luminosity ($\mathrm{L}_{CNO}$ and $\mathrm{L}_{pp}$) as a function of the central temperature $\mathrm{T}_\mathrm{c}$ for ZAMS models of different masses, computed adopting the $^{14}$N(p,$\gamma$)$^{15}$O from NACRE (\textit{red lines}) and LUNA (\textit{blue lines}) for the labelled chemical compositions.}
\label{fig:crossZ}
\end{figure}

Figure \ref{fig:crossZ} shows the ratio of the \textit{CNO} to $pp$ luminosities ($\mathrm{L}_{CNO}$ and $\mathrm{L}_{pp}$) versus the central temperature ($\mathrm{T}_{\mathrm{c}}$) for ZAMS models of different masses and metallicities computed adopting the two quoted cross-sections. For a given value of the central temperature, the $\mathrm{L}_{CNO}/\mathrm{L}_{pp}$ ratio drastically decreases as the metallicity becomes increasingly lower as a consequence of the lower C, N, and O abundances. Thus, the critical mass above which the effect of the adopted $^{14}$N(p,$\gamma$)$^{15}$O cross-section on the ZAMS location is not negligible increases with decreasing metallicity. As an example, this critical mass rises from about $1.5$ M$_\odot$ for $\mathrm{Z}=0.015$ to $\approx 2$~M$_\odot$ for $\mathrm{Z}=0.0002$. We note that in the plane ($\log \mathrm{T}_{\mathrm{c}}$, $\log \mathrm{L}_{CNO}/\mathrm{L}_{pp}$) the distance between the red and blue curves, computed respectively with NACRE and LUNA cross-section, is roughly independent of the metallicity.

Regarding the ZAMS position in the HR diagram, Fig.~\ref{fig:azoto}, the temperature differences decrease with metallicity: indeed for $\mathrm{Z}=0.0002$, we obtained $\Delta \mathrm{T}_{\mathrm{eff}}\approx30$ K for models of 2~M$_\odot$, 200~K for 3.0~M$_\odot$, and 300~K for 6~M$_\odot$.


\subsection{Initial conditions}
\label{sec:initial}
The choice of initial conditions, i.e. the starting model, is important when modelling the early phases of pre-MS evolution. The correct and physically consistent approach would be to take as a starting model the structure left at the end of the previous hydrodynamical evolution of the protostar, when the main mass- accretion process is finished. In this respect, as early as 1983 Stahler introduced the very useful concept of ``birthline'', defined as ``\textit{the locus in the HR diagram where pre-MS stars of various masses should first appear as visible objects}'' \citep{stahler83}. In agreement with this definition, realistic pre-MS models should start from the birthline, which should also play the role of a zero-age isochrone \citep{palla99}. 

However, the hydrodynamical protostellar evolution is still largely debated and not yet settled. Thus, we decided to follow the standard approach outlined by the pioneering contribution of \citet{Iben65} and followed until now by many authors \citep{BCAH98,SD00,yy01,dotter07,dicriscienzo09}, which consists of choosing as an initial model a fully convective structure of very large radius and luminosity, i.e. a point along the Hayashi track, from which the star begins its quasi-static contraction at constant mass, i.e. neglecting any mass accretion episode. 
The problem consists of understanding whether this simplistic assumption significantly affects the whole evolution of pre-MS models or only the very early stages. Many works agree in predicting that after the end of the main mass-accretion phase, the evolution quickly converge to that of standard hydrostatic models \citep{stahler83,palla91,palla92}. However, there is not yet complete agreement with this conclusion \citep{wuchterl03}.

\begin{figure}
\centering
\includegraphics[width=\linewidth]{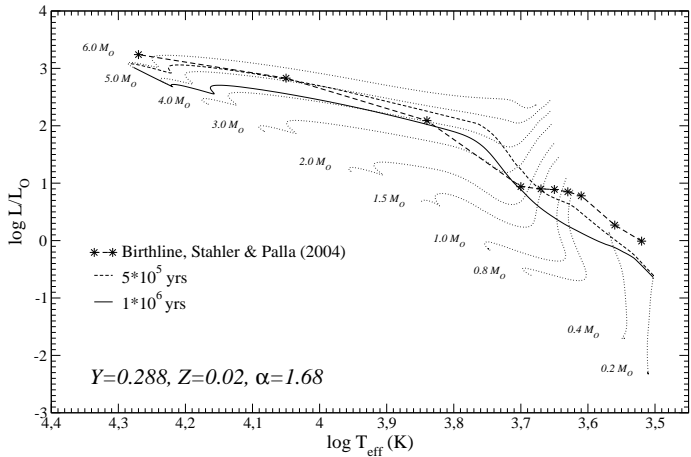}
\caption{Evolutionary tracks in the mass range 0.2$-$6.0 M$_\odot$ (\textit{dotted line}) with superimposed 0.5 Myr (\textit{dashed line}) and 1~Myr (\textit{solid line}) isochrones. The birthline by \citet{stahler04} is also shown (\textit{star-dashed line}).}
\label{fig:Birthline}
\end{figure}

We focus mainly on ages older than 1~Myr since the current theoretical predictions are generally considered unreliable for younger ages \citep[see e.g.,][]{tout99,siess01,baraffe02}. Figure \ref{fig:Birthline} shows evolutionary tracks ($0.2-6.0$~M$_\odot$) in the  HR diagram with the superimposed birthline of \citet{stahler04}, and our isochrones of 0.5 and 1~Myr. Standard stellar pre-MS models intersect the birthline at different ages, depending on the mass, but almost always before 1~Myr.
 
Another point that deserves to be discussed is the deuterium abundance of stars on the birthline. According to \citet{stahler04}, protostars start to burn deuterium once a total mass of 0.3~M$_\odot$ is reached, if a constant mass accretion rate of $10^{-5}$~M$_\odot\,\mathrm{yr}^{-1}$ is adopted. As a consequence, the initial deuterium abundance available on the birthline at the beginning of the pre-MS phase is exactly the value present in the ISM for stars of mass lower than 0.3~M$_\odot$, while it decreases significantly with increasing total mass. According to \citet{stahler04}, stars with a mass higher than about 1~M$_\odot$ should almost completely deplete the deuterium during their previous protostellar phase.

In contrast, in standard pre-MS models, the initial deuterium abundance is usually the same as that of the ISM. In principle, such a difference could affect the early pre-MS evolution, but in practice the effect is negligible, since in standard models the deuterium burning onset occurs at luminosities much higher than those corresponding to the birthline and when the model reaches it the deuterium depletion agrees with that provided by protostellar computations. As a test, we compared the evolution of a 0.9~M$_\odot$ model provided by our standard computation with that of \citet{stahler04}, which takes into account protostellar accretion. Starting from the same interstellar deuterium abundance, our model attains the luminosity of the birthline experiencing the same amount of deuterium depletion, namely a tenth of the interstellar value, as the model emerging from a protostellar evolution.  

The definition of proper starting model is also of interest because the age issue. Usually, the age assigned to pre-MS stars by means of standard models is simply the interval of time required for the model to attain the observed radius and luminosity starting from the arbitrary initial conditions of high luminosity and large radius. This  inferred age is inaccurate, since it completely neglects the protostellar phase, an inaccuracy that grows as the age of the star decreases. The correct approach should be to add the protostellar age to the pre-MS one, measured from the birthline \citep{stahler83}. 
On the other hand, this inaccuracy should be between 0.1 Myr and 0.6 Myr, the time needed to form a star according to \citet{stahler04} models which adopt a constant mass accretion rate of $10^{-5}$~M$_\odot\,\mathrm{yr}^{-1}$.

\section{Comparison with different authors}
\begin{table*}
\caption{\label{tab:inputs}Summary of the main characteristics of the selected grid of models. The columns provide the adopted EOS, radiative opacity (the heavy element mixture is specified), boundary condition characteristics, superadiabatic convection treatment, nuclear cross-sections and initial deuterium abundance. All these codes use a solar-calibrated convection efficiency, except DVD09.} 
\centering
\begin{tiny}
\begin{tabular}{c|ccccccc}    
\hline
\\
\textbf{Code: } & \textbf{EOS} & \textbf{Radiative opacity}  & \textbf{Boundary conditions} & \textbf{Convection} & \textbf{Reaction rates} & \textbf{X}$_{\mathrm{D}}$\\
 & & \textbf{(Mixture)}\\
\hline 
\hline
 &  & OPAL; & non-grey, $\tau_{\mathrm{ph}}=10$  & & NACRE; & $2\cdot 10^{-5}$, $\mathrm{Z}\ge 0.008$\\
 \textsc{\texttt{FRANEC}} & OPAL06  & F05 & \citet{brott} & MLT & LUNA &$4\cdot 10^{-5}$, $\mathrm{Z}<0.008$\\
 & & (AS05) & $\alpha_{\mathrm{atm}}=2$; & $\alpha=1.68$ & ($^{14}$N(p,$\gamma$)$^{15}$O) \\
 & & & \citet{castelli03}\\
 & & & $\alpha_{\mathrm{atm}}=1.25$\\
 \hline
\\
\textsc{BCAH98} & SCVH95  & OPAL;  & non-grey, $\tau_{\mathrm{ph}}=100$ & MLT & CF88 & $2\cdot 10^{-5}$ \\
&  & AF94   & \citet{allard97b} & $\alpha=1.9$\,\tablefootmark{a} \\
&  &  (GN93) & $\alpha_{\mathrm{atm}}=1$\\
\\
\textsc{DSEP08}  & \citet{chaboyer95}; & OPAL;   & non-grey, T$_{\tau_{\mathrm{ph}}}$=T$_{\mathrm{eff}}$ &  & AD98; & 0\\
&  \citet{irwin04} & F05  & \citet{hauschildt99,hauschildt99b} & MLT & LUNA04 \\
 & & (GS98)& $\alpha_{\mathrm{atm}}=2$; & $\alpha=1.938$ & ($^{14}$N(p,$\gamma$)$^{15}$O)\\
 & & &  \citet{castelli03} \\
 & &  & $\alpha_{\mathrm{atm}}=1.25$\\
 \\
\textsc{DVD09} & OPAL05; & OPAL;  & non-grey\,\tablefootmark{b}, M $<2$ M$_\odot$  \\
 &  SCVH95 & F05  & \citet{allard97b}, $\tau_{\mathrm{ph}}=3$ & MLT\,\tablefootmark{c} & NACRE & $4\cdot 10^{-5}$\\
 & & (GS99)   & $\alpha_{\mathrm{atm}}=1$; & $\alpha=2.0$;\\
 & & & \citet{heiter02}, FST, $\tau_{\mathrm{ph}}=10$; & FST \\
 & & & grey, M $\ge2$ M$_\odot$\\
 \\
\textsc{SD00} & modified version of & OPAL; & non-grey, $\tau_{\mathrm{ph}}=10$  & MLT & CF88 & $2\cdot 10^{-5}$\\
& PTEH & AF94  & \citet{kurucz91,plez92} & $\alpha=1.6$\\
& & (GN93) \\
\\
\textsc{YY01} & OPAL96; & OPAL; &  grey atmosphere, & MLT & BP92 & (\dots) \\
& \citet{chaboyer95} & AF94  & $\tau_{\mathrm{ph}}=2/3$ & $\alpha=1.743$\\
& & (GN93)\\
\hline 	
\hline
\end{tabular}
\tablefoot{\tablefoottext{a}{BCAH98 models have been computed also for $\alpha=1$; we used this latter value of $\alpha$ for the comparisons of 0.4 M$_{\,\,\odot}$.}
\tablefoottext{b}{DVD09 adopt the \citet{allard97} atmospheric models for M$\le 0.6\,\mathrm{M}_{\odot}$ and the \citet{heiter02} ones for M$> 0.6\,\mathrm{M}_{\odot}$.} 
\tablefoottext{c}{DVD09 models were computed adopting the MLT formalism for M$\le 0.6\,\mathrm{M}_{\odot}$ and the FST for M$> 0.6\,\mathrm{M}_{\odot}$.}}
\tablebib{
PTEH, \citet{pols95}; SCVH95, \citet{saumon95}; AD98, \citet{adelberger98}; BP92, \citet{bahcall92}; CF88, \citet{caughlan88}; LUNA04, \citet{imbriani04}; GS99, \citet{grevesse99}; AF94, \citet{alexander94}; FST, \textit{Full Spectrum of Turbulence} \citep{canuto91,canuto96}.
}
\end{tiny} 
\end{table*} 

We compare our reference set of tracks with some of the most recent and used pre-MS models available in the literature\footnote{The models selected for the comparisons obviously do not exhaust the sample of pre-MS tracks available in the literature, see e.g. \citet{DM97},  \citet{charbonnel99}, \citet{palla99}.}, namely, \citet[][BCAH98]{BCAH98}, \citet[][SD00]{SD00}, \citet[][YY01]{yy01}, \citet[][DSEP08]{dotter08}, and \citet[][DVD09]{dicriscienzo09}\footnote{DVD09, which is the updated version of \citet{DM97}, deal with low metallicities; the models with Z = 0.02 were computed by Marcella Di Criscienzo with the same evolutionary code described in DVD09.}.

\begin{figure*}
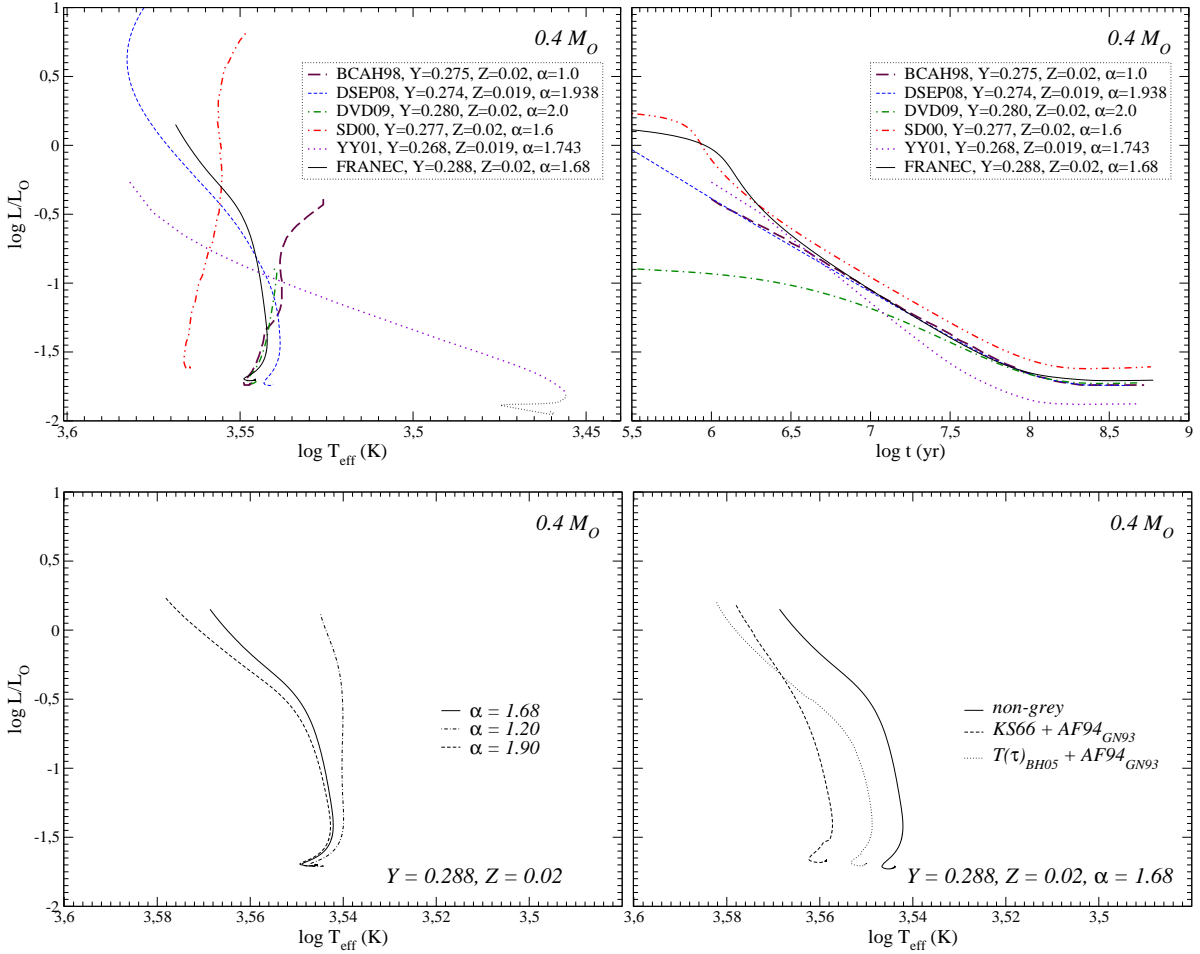

\centering
\includegraphics[width=0.85\linewidth]{Tutte_solar_calibrated_04_col_new.eps}
\\
\vspace{0.2cm}
\includegraphics[width=0.85\linewidth]{diffe_0p4_new.eps}
\caption{Upper panels: comparisons of 0.4 M$_\odot$ tracks from different authors listed in Table \ref{tab:inputs} for $\mathrm{Z} \approx 0.02$, in the HR diagram and in the ($\log  \mathrm{t\,(yr)}$, $\log \mathrm{L}/\mathrm{L}_\odot$) plane. Bottom left panel: 0.4 M$_\odot$ models computed with three values of the mixing length parameter, namely $\alpha=1.2,1.68$, and $1.9$. Bottom right panel: 0.4 M$_\odot$ models with three different choices of the adopted boundary conditions: non-grey models as described in Sect. \ref{sec:BC} (\textit{solid line}), KS66 T($\tau$) relation plus AF94 low-temperature opacity tables (\textit{dashed line}) and T($\tau$) relation interpolated from the \citet{brott} atmosphere models plus AF94 low-temperature opacity tables (\textit{dotted line}).}
\label{fig:diffe04}
\end{figure*}

\begin{figure*}
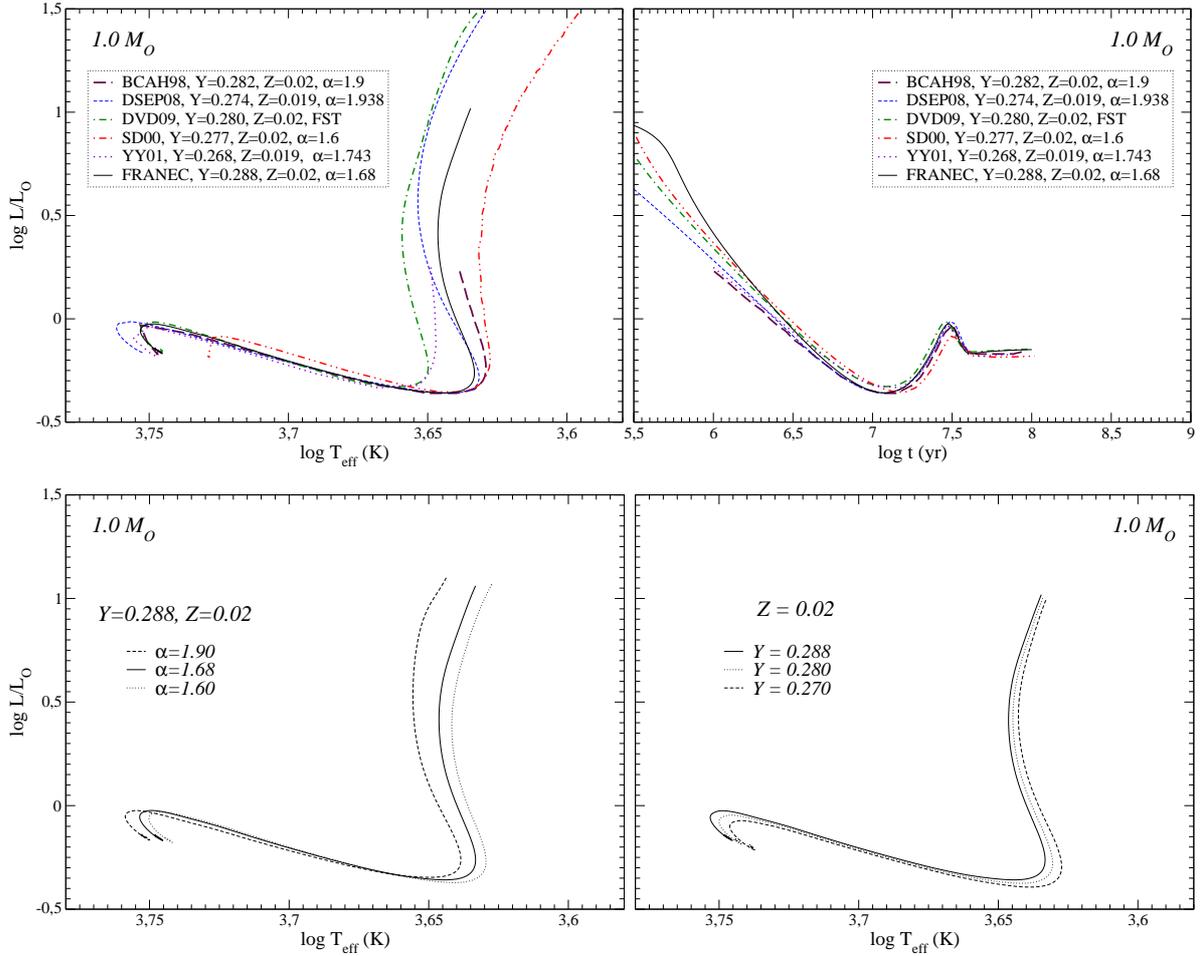

\centering
\includegraphics[width=0.85\linewidth]{Tutte_solar_calibrated_10_col_new.eps}
\\
\vspace{0.2cm}
\includegraphics[width=0.85\linewidth]{diffe_1p0.eps}
\caption{Upper panels: comparisons of 1 M$_\odot$ tracks from different authors listed in Table \ref{tab:inputs} with $\mathrm{Z} \approx 0.02$, in the HR diagram and in the ($\log  \mathrm{t\,(yr)}$, $\log \mathrm{L}/\mathrm{L}_\odot$) plane. Bottom left panel: 1 M$_\odot$ models computed with three values of $\alpha$, namely $\alpha = 1.6,\,1.68$, and 1.9 for $\mathrm{Z}= 0.02$ and $\mathrm{Y}=0.288$. Bottom right panel: 1 M$_\odot$ tracks with three different original helium abundances, $\mathrm{Y}=0.270$, $\mathrm{Y}=0.280$, and $\mathrm{Y}=0.288$ with $\mathrm{Z}=0.02$.}
\label{fig:diffe10}
\end{figure*}

\begin{figure*}
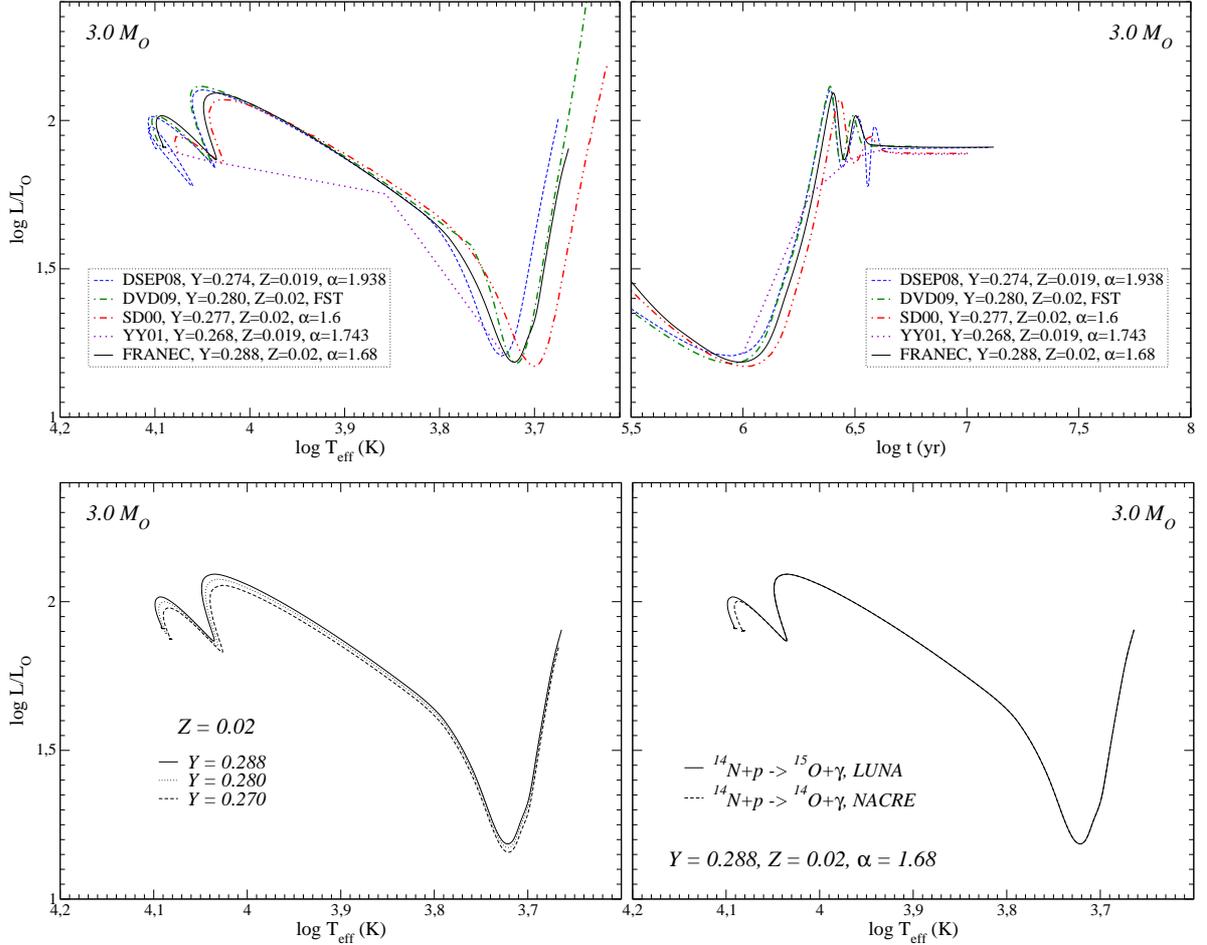

\centering
\includegraphics[width=0.85\linewidth]{Tutte_solar_calibrated_30_col_new.eps}
\\
\vspace{0.2cm}
\includegraphics[width=0.85\linewidth]{diffe_3p0.eps}
\caption{Upper panels: comparisons of 3 M$_\odot$ tracks from different authors listed in Table \ref{tab:inputs} for $\mathrm{Z} \approx 0.02$, in the HR diagram and in the ($\log \mathrm{t\,(yr)}$, $\log \mathrm{L}/\mathrm{L}_\odot$) plane. Bottom left panel: 3 M$_\odot$ models with three initial helium abundances, namely $\mathrm{Y}=0.270$, $\mathrm{Y}=0.280$, and $\mathrm{Y}=0.288$ with $\mathrm{Z} = 0.02$. Bottom right panel: 3 M$_\odot$ tracks computed adopting the LUNA $^{14}$N(p,$\gamma$)$^{15}$O reaction rate (\textit{solid line}) and the NACRE one (\textit{dashed line}).}
\label{fig:diffe30}
\end{figure*}

The upper panels of Fig. \ref{fig:diffe04}, \ref{fig:diffe10}, and \ref{fig:diffe30} show, respectively, the evolutionary tracks of 0.4 M$_\odot$, 1.0 M$_\odot$, and 3.0 M$_\odot$ with $\mathrm{Z}~\approx~0.02$. Table \ref{tab:inputs} lists the main characteristics (EOS, radiative opacity, boundary conditions, convection scheme, cross-sections, and initial deuterium abundance) of the quoted models.
\\
\\
\begin{figure*}
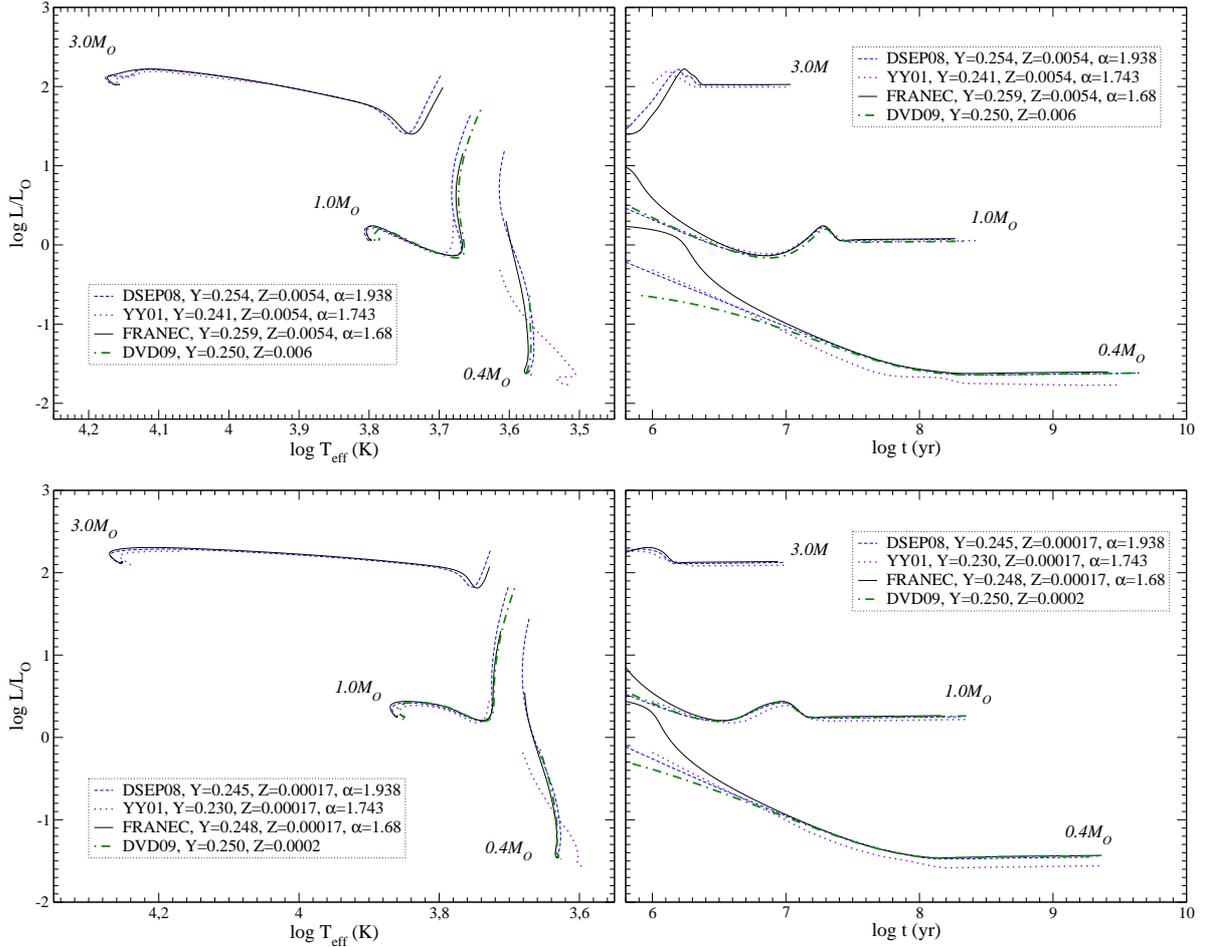

\centering
\includegraphics[width=0.85\linewidth]{Z6E3_DART_YY_DVD_col_new.eps}
\\
\vspace{0.2cm}
\includegraphics[width=0.85\linewidth]{Z2E4_DART_YY_DVD_col_new.eps}
\caption{Upper panels: comparisons between DSEP08, YY01, and \texttt{FRANEC} for $Z \approx 0.006$ in the HR diagram and in the ($\log \,\mathrm{t\,(yr)}$, $\log \mathrm{L/L}_\odot$) plane for 0.4 M$_\odot$, 1 M$_\odot$, and 3 M$_\odot$ models. Bottom panels: as in the upper panels but for a lower metallicity, namely  $\mathrm{Z}\approx 0.0002$.}
\label{fig:confronti_altri}
\end{figure*}
\noindent
\textbf{M = 0.4 M$_\odot$}
\\
\noindent
Figure \ref{fig:diffe04} (upper left panel) shows the HR diagram with the evolutionary tracks of 0.4 M$_\odot$. The models by BCAH98 with $\alpha=1$, DVD09 and \texttt{FRANEC} have a quite similar location in the HR diagram for $\log \mathrm{L/L}_\odot < -1$ as a result of using similar EOS and boundary conditions. Figure \ref{fig:diffe04} (bottom left panel) shows that the effect of adopting different $\alpha$  values becomes progressively negligible as the star approaches the ZAMS. More quantitatively, changing $\alpha$ from 1.68 to 1.2 produces a difference in effective temperature of about $150-170$~K for ages of 1 Myr, difference that decreases to 30 K at $\log \mathrm{L/L}_\odot = -1$ and to $10-15$ K in  ZAMS.

The model of DSEP08 is slightly colder (about 40 K) than those of BCAH98, DVD09, and \texttt{FRANEC} as it approaches the ZAMS; this difference is not easy to understand because the relevant input physics are similar. Moreover, we checked that the difference of $\alpha$ between DSEP08 and our models would produce a ZAMS hotter by about only 3 K (see bottom left panel of Fig. \ref{fig:diffe04}). The choice of $\tau_{\mathrm{ph}}$ could account, at least in part, for the observed discrepancy. DSEP08 uses $\tau_{\mathrm{ph}}$ such that T($\tau_{\mathrm{ph}}$)~=~$\mathrm{T}_{\mathrm{eff}}$, which leads to a variable value often lower than one. We verified that changing $\tau_{\mathrm{ph}}$ from 10 to the value such that T($\tau_{\mathrm{ph}}$)~=~$\mathrm{T}_{\mathrm{eff}}$ produces ZAMS about $15-20$~K colder. We also note that DSEP08 model has a lower original metal and helium abundances, but the two effects cancel each other.

The tracks of SD00 and YY01 show the greatest differences in both morphology and position on HR diagram compared to the others. SD00 use a modified version of the EOS described in \citet{pols95} that, as discussed in Sect. \ref{sec:eos}, should produce tracks colder than ours of about 35 K. A further shift (about $5-10$ K) towards lower effective temperatures due to the lower helium abundance is expected. However, Fig.~\ref{fig:diffe04} shows that the track is hotter than the other models for $\log \mathrm{L/L}_\odot <- 0.5$, reaching in ZAMS $\Delta \mathrm{T}_{\mathrm{eff}}\approx 150$ K. This result is not easy to explain. One might speculate that a not negligible role in the deviation of the SD00 track is played by the scheme adopted to integrate the atmosphere. As discussed in \citet{SD00}, the boundary conditions are obtained by integrating the T($\tau$) relationship resulting from a fit of atmospheric models. 

To test such a working hypothesis, we computed additional tracks (shown in the bottom right panel of Fig. \ref{fig:diffe04}), using two different T($\tau$) relationships, namely the classic KS66 (dashed line) and that obtained by interpolating the BH05 atmospheric models (dotted line), the latter representing the closest approximation to the SD00 choice. Both the models were computed using the same low-temperature radiative opacity adopted by SD00, i.e. AF94.  
The dotted line model is hotter than the reference one (solid line) of about $50-60$ K in ZAMS. Moreover, as discussed in Sect. \ref{sec:opacity}, the integration of a hydrostatic atmosphere for 0.4 M$_\odot$ stars is very sensitive to the extrapolation method used for the low-temperature opacity tables. This effect can contribute to increasing the differences in effective temperature.

For YY01, the track morphology is completely different from the others; the ZAMS is $500-550$ K colder than the BCAH98, DSEP08, DVD09, and \texttt{FRANEC} ones. This large discrepancy can be justified by neither the different initial helium abundance nor the adoption of the AF94 low-temperature radiative opacity in concomitance of the use of a grey T$(\tau)$ relationship (see bottom right panel of Fig. \ref{fig:diffe04}). It is not easy to unambiguously identify the reason for this behaviour without a full evolutionary computation with the same input physics. However, \citet{siess01} did prove that the location in the HR diagram of pre-MS tracks of low-mass stars is very sensitive to the contribution of molecular hydrogen in the EOS, showing that models computed with an EOS that does not account for H$_2$ are significantly less luminous ($\Delta \log \mathrm{L/L}_\odot \approx 0.6$ for a 0.3~M$_\odot$) and  colder (about 1000 K for a 0.3M$_\odot$) than those computed with an EOS that does take it into account. This seems to be the case for YY01, because for temperatures lower than 10$^6$ K, in the regions not covered by the OPAL EOS96, they adopt the Saha equation that accounts for a single state of ionization of atomic hydrogen and metals plus double states of ionization of helium but they neglect the contribution of H$_2$ \citep[see][]{guenther92}.

The right upper panel of Fig. \ref{fig:diffe04} shows the temporal evolution of the luminosity. The models by BCAH98, DSEP08, and \texttt{FRANEC} agree for $\log \mathrm{t\,(yr)}>~6.5-7$. The non-negligible discrepancy at earlier stages between the \texttt{FRANEC} and DSEP08 models ($\Delta \log \mathrm{L/L}_\odot\approx~0.3~-~0.4$ at $\log \mathrm{t\,(yr)} =6$) is caused by the different treatments of D-burning, DSEP08 assuming a zero initial deuterium abundance.

For $\log~\mathrm{t\,(yr)}~\approx~6$, BCAH98 is significantly fainter than SD00 and \texttt{FRANEC}, although these tracks adopt the same initial deuterium abundance. This behaviour can be related, at least in part, to the fact that BCAH98 track starts its evolution from an initial model with a central temperature high enough to ignite deuterium, at variance with SD00 and \texttt{FRANEC}, which start their evolution at much lower central temperature. \citet{dantona06} showed that the initial central temperature  significantly affects the D-burning phase, in particular tracks evolved from hotter initial models are fainter at a fixed age, as in the case of BCAH98. However, this effect is limited to a few Myr for a 0.4 M$_\odot$, which is the typical deuterium burning timescale.

The DVD09 model displays a quite peculiar behaviour that cannot be simply explained by the different initial deuterium abundance ($\mathrm{X}_{\mathrm{D}}=4\cdot  10^{-5}$). For $\log \mathrm{t\,(yr)} < 7.5$, it is systematically less luminous than the other models, with a maximum discrepancy in luminosity of about $\Delta \log\mathrm{L/L}_\odot \approx 0.9$ at 1 Myr. One might speculate that such a large difference in luminosity for such a low-mass star is related to the D-burning phase at the beginning of the Hayashi track that is not included in DVD09 computations. \citet{dicriscienzo09} discussed how, owing to the lack of atmospheric structures for $\log \mathrm{g\,(cm\,s^{-2})} < 3.5$, they had to skip the deuterium burning during the early evolutionary phases. The DVD09 0.4 M$_\odot$ track begins at $\log \mathrm{g\,(cm\,s^{-2})} \approx 4.00$, at a luminosity lower than the other models ($\log \mathrm{L/L}_\odot \approx -1$), BCAH98 included, who use the same atmospheric models. As the star approaches the ZAMS the DVD09 model eventually converges to BCAH98.

As already seen in the HR diagram, the YY01 model shows a temporal evolution of luminosity quite different from the other ones. While the other models converge after the D-burning phase, the YY01 one becomes progressively less luminous until it reaches the ZAMS with a maximum difference of about $\Delta \log \mathrm{L/L}_\odot \approx~0.2$.
\\
\\
\noindent
\textbf{M  = 1.0 M$_\odot$}
\\
\noindent
Figure \ref{fig:diffe10} compares the tracks of 1 M$_\odot$. All the models predict almost the same luminosity for the ZAMS. The effective temperatures of the models of BCAH98, DVD09, YY01, and \texttt{FRANEC} are still in reciprocal good agreement, whereas DSEP08 and SD00 ZAMS are, respectively, hotter by about 94 K and colder by about 210 K than ours.

As previously shown in Sect. \ref{sec:eos}, the ZAMS position of 1 M$_\odot$ stars in the HR diagram is quite insensitive to the EOS, i.e. the difference between PTEH and OPAL models is about 15 K. In contrast, as shown in Sect. \ref{sec:opacity}, the chosen heavy-element mixture has a significant effect on the ZAMS, in particular substituting the old GN93 with the most recent AS05 mixture leads to colder ($\approx 90$ K) and fainter ($\approx 0.05$) 1 M$_\odot$ models. DSEP08 adopt GS98 and therefore the difference should be smaller. However, for DSEP08 this shift in the effective temperature is largely balanced by the different boundary conditions chosen. We verified that when we adopt the same prescription as DSEP08 (T($\tau_{\mathrm{ph}}$) = T$_{\mathrm{eff}}$) the position of the ZAMS is shifted towards lower temperatures by about 50 K with respect to the model computed with $\tau_{\mathrm{ph}} = 10$. Moreover, the DSEP08 model adopts $\alpha\approx 1.9$, which would produce a ZAMS hotter than that computed with $\alpha=1.68$ for about 70 K, as shown in the bottom left panel of Fig. \ref{fig:diffe10}. Therefore, the net effect of the different heavy-element mixture, boundary conditions, and $\alpha$ values between the \texttt{FRANEC} and DSEP08 models is to produce a difference in the effective temperature on the ZAMS of the order of 100 K, which is fully compatible with the 94 K actually present. The slight difference between the initial metal and helium abundances of DSEP08 and our model is inconsequential because the effect of increasing helium is counterbalanced by the concomitant reduction in metals.

The track of SD00 is computed with $\alpha=1.6$, which should produce a ZAMS colder than ours by about 40 K (bottom left panel of Fig. \ref{fig:diffe10}). In this case, the total effect of the different heavy-element mixture and mixing length parameter would be a ZAMS model hotter than the \texttt{FRANEC} one of about 60~K, but actually  colder of 210 K. This large discrepancy can not be entirely justified by the different helium abundance. As shown in the bottom right panel of Fig. \ref{fig:diffe10}, even a difference larger than that between the \texttt{FRANEC} and SD00 models, namely $\Delta \mathrm{Y}=-0.018$, would reduce the ZAMS effective temperature by about 100 K making SD00 colder than our track of about 40 K. However, the peculiar behaviour of the SD00 1 M$_\odot$ model with $\mathrm{Z}=0.02$ was previously discussed by \citet{montalban04b}. 

The maximum difference in temperature along the Hayashi track reaches almost 300 K between DVD09, which is the hottest model, and SD00, the coolest. The agreement among the other tracks is better, i.e. $\Delta \mathrm{T}_{\mathrm{eff}}\approx$ 60 K between \texttt{FRANEC} and BCAH98. With the exceptions of DVD09 and YY01, the pre-MS models tend to converge at the end of the Hayashi track when stars move towards the ZAMS. We note that the lower the helium abundance the colder the Hayashi track, with a shift in effective temperature of about 36 K for $\Delta \mathrm{Y} = 0.018$ at $\log \mathrm{L/L}_\odot = 0.5$ (see Fig. \ref{fig:diffe10} bottom right panel). However, the dominant role in determining the position of the Hayashi track is played by the efficiency of the superadiabatic convection. Figure \ref{fig:diffe10} bottom left panel shows the effect of changing the $\alpha$ value between 1.9 and 1.6 in the interior ($\Delta \mathrm{T}_{\mathrm{eff}}\approx 150$~K at $\log \mathrm{L/L}_\odot = 0.5$). 

The upper right panel in Fig. \ref{fig:diffe10} shows the time evolution of luminosity. At the age of 1 Myr, the maximum difference between the models is $\Delta \log \mathrm{L/L}_\odot \approx 0.2$ and decreases as the models evolve, reaching $\Delta \log \mathrm{L/L}_\odot \approx 0.03-0.04$ in ZAMS.
\\
\\
\noindent
\textbf{M = 3.0 M$_\odot$}
\\
\noindent
Figure \ref{fig:diffe30} shows the evolution of a 3 M$_\odot$. The models by BCAH98 are unavailable for M $> 1.4$ M$_\odot$, while the track of YY01 has only a few points making the comparison quite difficult and not so meaningful. Thus, we present only the comparison with DSEP08, DVD09, and SD00 tracks. Moreover, we do not discuss the location of the Hayashi track, since the star takes about 1 Myr to leave it, but focus instead on describing the discrepancies between the models for two specific evolutionary phases, namely the first relative maximum in effective temperature before the steep drop in luminosity (hereafter hook\footnote{This phase corresponds approximately to the first model which is completely supported by central hydrogen burning with the secondary elements not yet being in equilibrium.}) and the ZAMS. We note that these two points are insensitive to the efficiency of the superadiabatic convection owing to the lack of a significant convective envelope.

The discrepancy in the hook effective temperature between the models is about 310 K for DSEP08 and \texttt{FRANEC}, 360 K for DVD09 and \texttt{FRANEC}, and about -150 K for SD00 and \texttt{FRANEC}. To help us understand the origin of these differences, the bottom panels of Fig. \ref{fig:diffe30} show models computed with three values of Y, namely $0.288$, $0.280$, and $0.270$ for $\mathrm{Z}=0.02$ (bottom left panel), and with two values of the $^{14}$N(p,$\gamma$)$^{15}$O nuclear cross-section (bottom right panel), at fixed chemical composition, one by the NACRE compilation and the other by the latest release of the LUNA facility. The original helium abundance variation accounts for differences with respect to the \texttt{FRANEC} model of about -150 K for DSEP08 and -100~K for DVD09, whereas the effect of the $^{14}$N(p,$\gamma$)$^{15}$O reaction rate becomes important only near the ZAMS and is completely negligible in the hook. We note that DSEP08 is also slightly metal poorer than the other models: this leads to a shift of about 110 K towards higher effective temperatures, hence the effect of the lower initial helium abundance is approximatively counterbalanced by the lower metal content. We recall that the position of the hook is also quite sensitive to the heavy-element mixture, that, as discussed in Sect. \ref{sec:mixture}, would account for a difference of about 300 K in effective temperature and about 0.03 in $\log \mathrm{L/L}_\odot$ between tracks computed with the GN93 and AS05 mixtures. A small correction to this estimate is expected, since DSEP08 and DVD09 use a slightly different heavy-element mixture. Thus, because of the effects of the adopted chemical composition and solar mixture, we would expect a net shift towards higher effective temperatures than our model of about 260 K for DSEP08 and 200 K for the DVD09, which is not enough to justify the observed differences of, respectively, 310 K and 360 K.

At the ZAMS location, we found discrepancies in effective temperature with respect to \texttt{FRANEC} of about 200 K for DSEP08, 100 K for DVD09, and -400 K for SD00. The effect of the different helium abundances and heavy element mixtures are similar to those mentioned above. In addition, as discussed in Sect.~\ref{sec:deuterium} and shown in the right bottom panel of Fig. \ref{fig:diffe30}, the ZAMS location is very sensitive to the adopted $^{14}$N(p,$\gamma$)$^{15}$O reaction rate. The difference in effective temperature between tracks computed with the NACRE and LUNA cross-section is about -230 K, a value that should also be representative of the difference between LUNA and \citet{caughlan88} reaction rates used by SD00 \citep[see e.g.,][]{nacre}. Owing to the different initial helium abundances and heavy-element mixtures adopted, we expect the DVD09 and SD00 models to be respectively 30 K and 60 K colder than the \texttt{FRANEC} one, while DVD09 in contrast is 100 K hotter and SD00 is about 340 K colder than our prediction.

DSEP08 use the old LUNA $^{14}$N(p,$\gamma$)$^{15}$O reaction rate \citep{imbriani04}, which is about $6\%$ greater than the latest LUNA release adopted in \texttt{FRANEC}. This difference would make the DSEP08 model colder than ours by about 20 K. Adding to this the other differences in the physical inputs adopted by DSEP08 with respect to \texttt{FRANEC} (i.e. initial helium and metal abundances and mixture) would account for a discrepancy of about 240 K, which is approximatively 40 K higher than what is observed.

The pre-MS time evolution of the luminosity for this mass is shown in the right upper panel of Fig. \ref{fig:diffe30}. Before the hook ($6<\log \mathrm{t\,(yr)} <6.4$),  \texttt{FRANEC} and SD00 models, for a fixed age, are less luminous than DVD09 ($\Delta \log \mathrm{L/L}_\odot\approx 0.07$ and $\Delta \log \mathrm{L/L}_\odot \approx 0.1$ respectively). The ZAMS age of DVD09 and \texttt{FRANEC} are in reciprocal good agreement ( $\Delta \mathrm{t}/\mathrm{t} \la 5\%$), whereas SD00 predict a ZAMS age older by about $20-30\%$ than \texttt{FRANEC} and DVD09.

In addition the luminosity evolution as a function of age of DSEP08 is in good agreement with our model, but  we note that there is a peculiar bump near the ZAMS, which  none of the other models show.

\subsection{Comparison between metal poor models.}
Figure \ref{fig:confronti_altri} shows the comparison for sub-solar metallicity, namely $\mathrm{Z} \approx 0.006$ (top panels) and $\mathrm{Z}\approx 0.0002$ (bottom panels), between our models and those of DVD09, DSEP08, and YY01\footnote{BCAH98 and SD00 metal-poor models are unavailable. We also note that DVD09 low-metallicity models are available only for M$\le 1.5$ M$_\odot$.}. We note that although the initial metallicities are very similar, if not identical, the initial helium abundances are quite different. Thus, the observed discrepancies in the HR diagram location, in particular near the ZAMS, are not only the result of different physical inputs but also of the adopted initial chemical compositions. However, the models are in reasonable agreement with the exception of the YY01 0.4 M$_\odot$ track, which is significantly colder than the others (by more than 300 K in ZAMS), probably as a consequence of the adopted EOS, as previously discussed. We note that, at variance with the case for  $\mathrm{Z}=0.02$, the effect of adopting different heavy-element mixtures is small for models with $\mathrm{Z}\approx 0.006$ and completely negligible in the case of $\mathrm{Z} \approx 0.0002$, as we discussed at the end of Sect. \ref{sec:mixture}. 

The agreement in the HR diagram is very good between \texttt{FRANEC} and DSEP08 across the whole range of masses for both the metallicities, with the exception of 0.4 M$_\odot$ which is slightly colder than ours by about $40-50$ K. A quite good agreement is achieved also between \texttt{FRANEC} and YY01 for $\mathrm{Z} =0.0054$ (0.4~M$_\odot$ excluded), whereas the $\mathrm{Z} = 0.00017 $ YY01 tracks have a ZAMS that is colder than ours by about $3-4 \%$. When comparing DVD09 and \texttt{FRANEC}, very small differences are present near the Hayashi track for both the metallicities, which increase near the ZAMS of 1.0 M$_\odot$ where DVD09 models are colder than DSEP08 and \texttt{FRANEC} by about $130-150$ K.

As one can see in the right panels of Fig. \ref{fig:confronti_altri}, the temporal evolution of the luminosity for metal-poor models displays the same features as discussed for the $Z\approx 0.02$ case.

\section{Comparison with observations}
The above discussion confirms a well-known feature of pre-MS tracks and isochrones: their position in the HR diagram is very sensitive to many still poorly constrained factors. The corresponding uncertainty in the predicted $\mathrm{T}_{\mathrm{eff}}$ and L directly propagates into an uncertainty in the inferred masses and ages of observed young stars and consequently in the IMF and SFR of young stellar systems. Hence, it is of crucial importance to compare the theoretical predictions with  observations. 

\begin{figure}
\centering
\includegraphics[width=\linewidth]{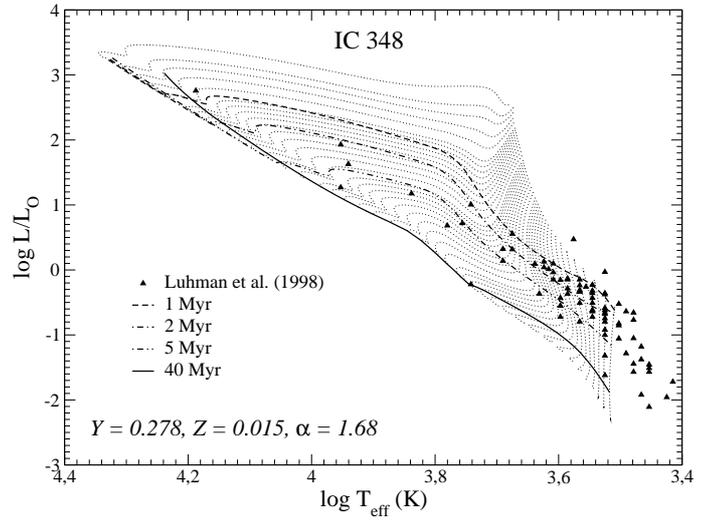}
\caption{HR diagram of the core of the cluster IC 348 by \citet{luhman98} with our theoretical pre-MS track superimposed for the mass range  0.2 M$_\odot$ to 7.0 M$_\odot$ (\emph{dotted lines}) and isochrones of ages 1 Myr (\emph{dashed line}), 2 Myr (\emph{dot-dashed line}), 5 Myr (\emph{dot-dot-dashed line}), and 40 Myr (\emph{solid line}).}
\label{fig:Luhman}
\end{figure}

Figure \ref{fig:Luhman} shows the HR diagram of the core of the cluster IC~348 studied by \citet{luhman98} with our theoretical pre-MS tracks and isochrones superimposed. With the exception of the faintest tail, the agreement is quite good.

\begin{figure}
\centering
\includegraphics[width=\linewidth]{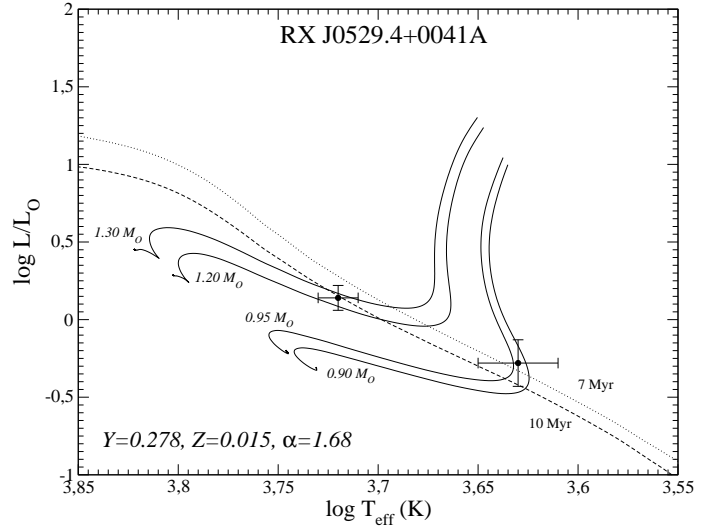}
\caption{HR diagram of the binary system RX J0529.4+0041A \citep{covino04} with our theoretical pre-MS tracks of labelled mass (\emph{solid lines})
and isochrones of 7 Myr (\emph{dotted line}) and 10 Myr (\emph{dashed line}) superimposed.}
\label{fig:RX}
\end{figure}

However, a much more severe test of pre-MS models is binary stars, especially detached, double-lined, eclipsing systems, which allow us to calibrate the theoretical tracks against stars of known mass. Figure \ref{fig:RX} shows the comparison between the RX~J0529.4+0041A binary system \citep{covino00,covino04} and the present theoretical tracks and isochrones. This  binary has been extensively studied \citep[see e.g.,][]{dantona01,baraffe02,montalban04b,montalban06} and provides a very useful and tight benchmark for pre-MS evolutionary tracks, because the two low-mass components (M$_1$=~1.27~$\pm$~0.01~M$_\odot$ and M$_2$=~0.93~$\pm$~0.01~M$_\odot$) are still in the pre-MS phase. Even if an exhaustive empirical test of the tracks should rely on a detailed discussion of the effects of the various sources of uncertainties, not least the chemical composition, we can safely claim that the agreement is quite good and very encouraging.
 
We note that among the current theories of binary formation there is not yet consensus on the coevality of the two stellar components, thus we should not in principle expect the two components of a binary system to have the same age \citep[see e.g. comment in][]{palla01}. This means that we should not pretend that both the stars of a binary are fitted by the same isochrone. On the other hand, the comparison between theory and observation seems to suggest coevality, that is the components of binaries form at about the same age \citep[see e.g.][]{hartigan94,brandner97,palla01}.

\section{The database}
\label{sec:database}
\begin{figure}
\centering
\includegraphics[width=\linewidth]{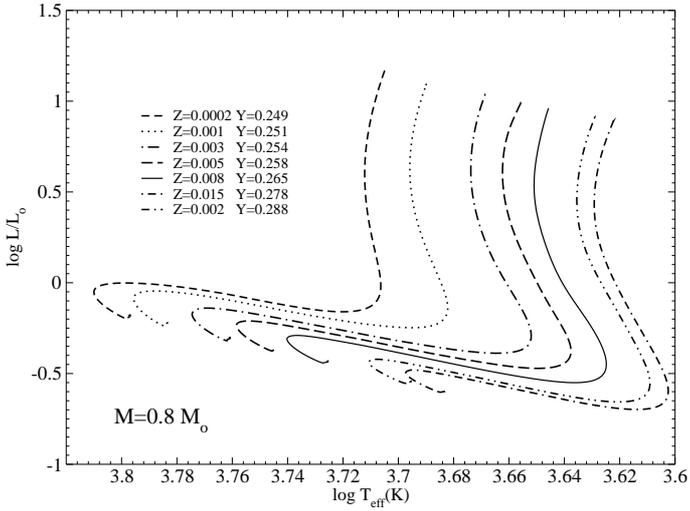}
\caption{HR diagram with 0.8 M$_\odot$ pre-MS tracks with the different labelled chemical compositions.}
\label{fig:chimica_tot}
\end{figure}

\begin{figure}
\centering
\includegraphics[width=\linewidth]{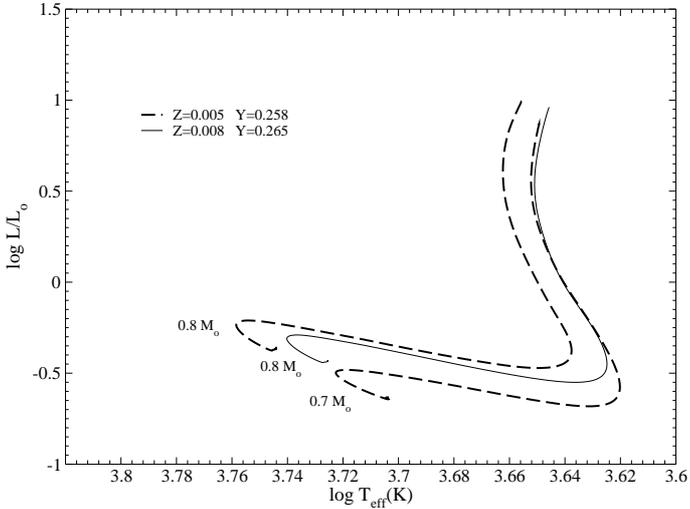}
\caption{Pre-MS tracks of 0.8 and 0.7 M$_\odot$ with $\mathrm{Z}=0.005$ and $\mathrm{Y}~=~0.258$ (\emph{dashed lines}) and of 0.8 M$_\odot$ with $\mathrm{Z}=0.008$ and $\mathrm{Y}=0.265$  (\emph{solid line}).}
\label{fig:chimica2}
\end{figure}

As is well known, the location in the HR diagram of pre-MS tracks, in particular the Hayashi line, strongly depends on the chemical composition, and particularly the metallicity. As an example of how large this effect can be, Fig. \ref{fig:chimica_tot} shows 0.8 M$_\odot$ models with different chemical compositions. Figure \ref{fig:chimica2} shows the results for two selected metallicities, namely $\mathrm{Z}=0.005$ and 0.008, corresponding to a difference in [Fe/H] of about 0.2 dex, roughly twice as much as the typical spectroscopic uncertainty. The shift in effective temperature of the Hayashi track at fixed mass is about 100~K. This means that to reproduce the Hayashi track location of a model with M=~0.8~M$_\odot$ and $\mathrm{Z} = 0.008$ with tracks of $\mathrm{Z}=0.005$, a decrease in the mass of 0.1 M$_\odot$ is required. Furthermore, this metallicity error also affects the age, since the contraction timescale of a 0.7 M$_\odot$ star is longer than that of a 0.8~M$_\odot$ star\footnote{At $\log \mathrm{L/L}_\odot$=-0.4, where the model with $\mathrm{Z}=0.008$ starts to leave the Hayashi track, the age of the 0.7 M$_\odot$ model is $70\%$ older than the one of 0.8 M$_\odot$ with $\mathrm{Z}=0.008$.}. This simple example gives an idea of the error in the inferred mass and age of stars in the pre-MS phase caused by an error in the assumed chemical composition. Thus, when comparing data with theoretical pre-MS tracks, one should be very careful to use models with the same chemical composition (in particular the same metallicity) of the observed stars.

This is why we provide a database of models with a very fine grid of Z and Y values. The present models are available for 19 metallicity values, from $\mathrm{Z}=0.0002$ to $\mathrm{Z}=0.03$. For each value of Z, we computed models for three different initial helium abundances Y, following the relation
\begin{equation}
\mathrm{Y} = \mathrm{Y}_{\mathrm{p}} + \frac{\Delta \mathrm{Y}}{\Delta \textrm{Z}}\textrm{Z},
\end{equation} 
where $\mathrm{Y}_{\mathrm{p}}$ represents the cosmological $^4$He abundance and $\Delta \mathrm{Y}$/$\Delta \mathrm{Z}$ is the Galactic helium-to-metal enrichment ratio. For the cosmological value of $\mathrm{Y}_{\mathrm{p}}$, we used both the recent WMAP estimate $\mathrm{Y}_{\mathrm{p}}=0.2485$ \citep{cyburt04,steigman06,peimbert07b,peimbert07a} and an older estimate of $\mathrm{Y}_{\mathrm{p}}=0.230$ \citep{lequeux79,pagel89,olive91}, usually referred to as canonical $\mathrm{Y}_{\mathrm{p}}$ in several stellar isochrone databases \citep[e.g, ][]{vandenberg00,girardi00,yy01,girardi02,cariulo04}. 

The estimated value of helium to metal enrichment is affected by several uncertainties, thus we decided to choose both the typical value $\Delta \mathrm{Y}$/$\Delta \mathrm{Z}=2$ and a higher value $\Delta \mathrm{Y}$/$\Delta \mathrm{Z}=5$, which seems to be the upper extreme 
\citep{pagel98,jimenez03,flynn04,gennaro10}. For each value of $\mathrm{Y}_{\mathrm{p}}$, $\Delta \mathrm{Y}$/$\Delta \mathrm{Z}$, and Z, we computed models in the mass range 0.2$-$7.0~M$_\odot$ with different mass steps: 0.05~M$_\odot$ in the mass range 0.2$-$1.0~M$_\odot$, 0.1~M$_\odot$ in the range 1.0$-$2.0~M$_\odot$, 0.2~M$_\odot$ for the mass range 2.0$-$4.0~M$_\odot$, and 0.5~M$_\odot$ in the range 4.0$-$7.0~M$_\odot$. 

\begin{table*}[htbp]
\caption{Summary of the models and isochrones available in the database. The table lists the initial deuterium abundance $\mathrm{X}_{\mathrm{D}}$, the mixing length parameter ($\alpha$), the primordial abundance of helium ($\mathrm{Y}_{\mathrm{p}}$), the helium to metal enrichment ratio ($\Delta\mathrm{Y}/\Delta \mathrm{Z}$), the initial helium (Y), and metal (Z) abundance.}
\label{tab:database}
\large
\centering
\begin{tabular}{|c||c|c|c||c|c|c||c|c|c|c||}     
\multicolumn{1}{c}{} &\multicolumn{6}{c}{\textrm{X$_{\mathrm{D}}=4.0\cdot 10^{-5}$}} & \multicolumn{1}{c}{} & \multicolumn{3}{c}{\textrm{X$_{\mathrm{D}}=2.0\cdot 10^{-5}$}}\\
\cline{2-7} \cline{9-11}
\multicolumn{1}{c}{} & \multicolumn{3}{|c||}{\textrm{$\alpha = 1.68$}} & \multicolumn{3}{c||}{\textrm{$\alpha = 1.2,\,1.9$}} & \multicolumn{1}{c}{} & \multicolumn{3}{|c||}{\textrm{$\alpha=1.2,\,1.68,\,1.9$}}\\
\cline{1-7} \cline{9-11}
 \multicolumn{1}{|c||}{\textrm{Y$_{\mathrm{p}}$ = }} & \multicolumn{1}{c}{0.230} & \multicolumn{2}{|c||}{0.2485} & \multicolumn{1}{c}{0.230} & \multicolumn{2}{|c||}{0.2485} & \multicolumn{1}{c}{} & \multicolumn{1}{|c}{\textrm{0.230}} & \multicolumn{2}{|c||}{\textrm{0.2485}} \\
\cline{1-7} \cline{9-11}
$\Delta$Y/$\Delta$Z = & 2 & 2 & 5 & 2 & 2 & 5 & \multicolumn{1}{c|}{} & 2 & 2 & 5\\
\hline
\hline
 Z: & \multicolumn{3}{c||}{ Y: } & \multicolumn{3}{c||}{ Y: } &  \multicolumn{1}{c|}{} & \multicolumn{3}{c||}{ Y: }\\
 \cline{1-7} \cline{9-11}
 $2.00\cdot 10^{-4}$ & 0.230 & 0.249 & 0.250 & 0.230 & 0.249 & 0.250 & \multicolumn{1}{|c|}{} &   & & \\
\cline{1-7} \cline{9-11}
 $1.00\cdot 10^{-3}$ & 0.232 & 0.251 & 0.254 & 0.232 & 0.251 & 0.254 & \multicolumn{1}{c|}{} &  &  & \\
\cline{1-7} \cline{9-11}
 $2.00\cdot 10^{-3}$ & 0.234 & 0.253 & 0.259 & 0.234 & 0.253 & 0.259 & \multicolumn{1}{c|}{} &  &  & \\
\cline{1-7} \cline{9-11}
 $3.00\cdot 10^{-3}$ & 0.236 & 0.254 & 0.263 & 0.236 & 0.254 & 0.263 & \multicolumn{1}{c|}{} &  &  & \\
\cline{1-7} \cline{9-11}
 $4.00\cdot 10^{-3}$ & 0.238 & 0.256 & 0.269 & 0.238 & 0.256 & 0.269 & \multicolumn{1}{c|}{} &  &  & \\
\cline{1-7} \cline{9-11}
$5.00\cdot 10^{-3}$ & 0.240 & 0.258 & 0.273 & 0.240 & 0.258 & 0.273 &  \multicolumn{1}{c|}{} &  &  & \\
\cline{1-7} \cline{9-11}
 $6.00\cdot 10^{-3}$ & 0.242 & 0.260 & 0.279 & 0.242 & 0.260 & 0.279 &  \multicolumn{1}{c|}{} &  &  & \\
\cline{1-7} \cline{9-11}
 $7.00\cdot 10^{-3}$ & 0.244 & 0.262 & 0.283 & 0.244 & 0.262 & 0.283 &  \multicolumn{1}{c|}{} &  &  & \\
\cline{1-7} \cline{9-11}
 $8.00\cdot 10^{-3}$ & 0.246  & 0.265 & 0.289 & 0.246 & 0.265 & 0.289 & \multicolumn{1}{c|}{} & 0.246  &  0.265 & 0.289\\
\cline{1-7} \cline{9-11}
 $9.00\cdot 10^{-3}$ & 0.248  & 0.267 & 0.294 & 0.248 & 0.267 & 0.294 & \multicolumn{1}{c|}{} & 0.248  &  0.267 & 0.294\\
\cline{1-7} \cline{9-11}
 $1.00\cdot 10^{-2}$ & 0.250 & 0.268 & 0.299 & 0.250 & 0.268 & 0.299 & \multicolumn{1}{c|}{} & 0.250 &  0.268 & 0.299 \\
\cline{1-7} \cline{9-11}
 $1.25\cdot 10^{-2}$ & 0.255 & 0.274 & 0.311 & 0.255 & 0.274 & 0.311 & \multicolumn{1}{c|}{} & 0.255 &  0.274 & 0.311 \\
\cline{1-7} \cline{9-11}
 $1.50\cdot 10^{-2}$ & 0.260 & 0.278 & 0.323 & 0.260 & 0.278 & 0.323 & \multicolumn{1}{c|}{} & 0.260 &  0.278 & 0.323 \\
\cline{1-7} \cline{9-11}
 $1.75\cdot 10^{-2}$ & 0.265 & 0.284 & 0.336 & 0.265 & 0.284 & 0.336 & \multicolumn{1}{c|}{} & 0.265 &  0.284 & 0.336 \\
\cline{1-7} \cline{9-11}
 $2.00\cdot 10^{-2}$ & 0.270 & 0.288 & 0.349 & 0.270 & 0.288 & 0.349  & \multicolumn{1}{c|}{} & 0.270 &  0.288 & 0.349 \\
\cline{1-7} \cline{9-11}
 $2.25\cdot 10^{-2}$ & 0.275 & 0.294 & 0.361 & 0.275 & 0.294 & 0.361 & \multicolumn{1}{c|}{} & 0.275 &  0.294 & 0.361 \\
\cline{1-7} \cline{9-11}
 $2.50\cdot 10^{-2}$ & 0.280 & 0.299 & 0.374 & 0.280 & 0.299 & 0.374 & \multicolumn{1}{c|}{} & 0.280 &  0.299 & 0.374 \\
\cline{1-7} \cline{9-11}
 $2.75\cdot 10^{-2}$ & 0.285 & 0.304 & 0.386 & 0.285 & 0.304 & 0.386 & \multicolumn{1}{c|}{} & 0.285 &  0.304 & 0.386 \\
\cline{1-7} \cline{9-11}
 $3.00\cdot 10^{-2}$ & 0.290 &  0.308 & 0.398 &0.290 & 0.308  & 0.398 &  \multicolumn{1}{c|}{} & 0.290 & 0.308 & 0.398 \\
\cline{1-7} \cline{9-11}
\multicolumn{11}{c}{}\\
\cline{1-11}
\multicolumn{3}{|c|}{\textrm{Solar chemical composition}} & \multicolumn{2}{c|}{\textrm{Y = 0.2533}} & \multicolumn{2}{c|}{\textrm{Z = 0.01377}} & \multicolumn{2}{c|}{\textrm{$\alpha=1.68$}} & \multicolumn{2}{c|}{\textrm{X$_{\mathrm{D}}=2.0\cdot 10^{-5}$}}\\
\cline{1-11}
\end{tabular}     
\end{table*}

The models are available for three different values of the mixing length parameter $\alpha$, namely 1.68 (solar calibrated), 1.2, and 1.9.
 
All models adopt an initial deuterium abundance $\mathrm{X}_{\mathrm{D}}=~4~\cdot~10^{-5}$, which is a good estimation at least for population II stars. As discussed previously in Sect. \ref{sec:deuterium}, population I stars should have a lower deuterium abundances, hence models with X$_{\mathrm{D}}=2\cdot 10^{-5}$ for $\mathrm{Z}\ge 0.008$ are also available.

The grid of models available in the database are summarized in Table \ref{tab:database}. We also provide an additional set of models corresponding to the chemical composition of our SSM, namely $\mathrm{Z}=0.01377$ and $\mathrm{Y}=0.2533$. 

\begin{figure*}
\centering
\includegraphics[width=\linewidth]{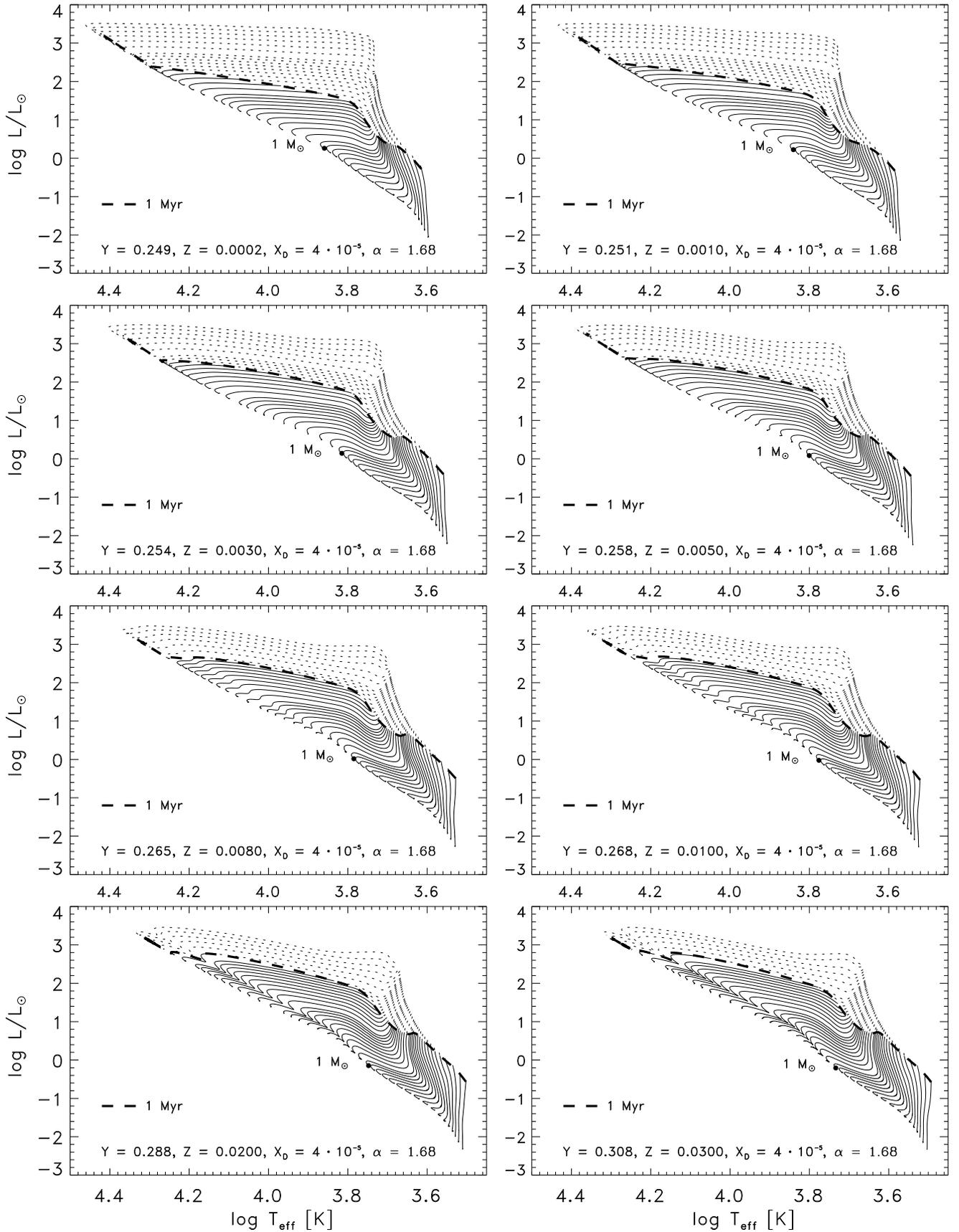}
\caption{Theoretical tracks in the mass range 0.2 $-$ 7 M$_\odot$ for labelled chemical composition with $\alpha=1.68$ and X$_{\mathrm{D}}=4\cdot 10^{-5}$ with superimposed the corresponding isochrone of 1~Myr (\textit{dashed line}). The dotted line represents the evolutionary phases with ages younger than 1~Myr.}
\label{fig:multi_track}
\end{figure*}

\begin{figure*}
\centering
\includegraphics[width=\linewidth]{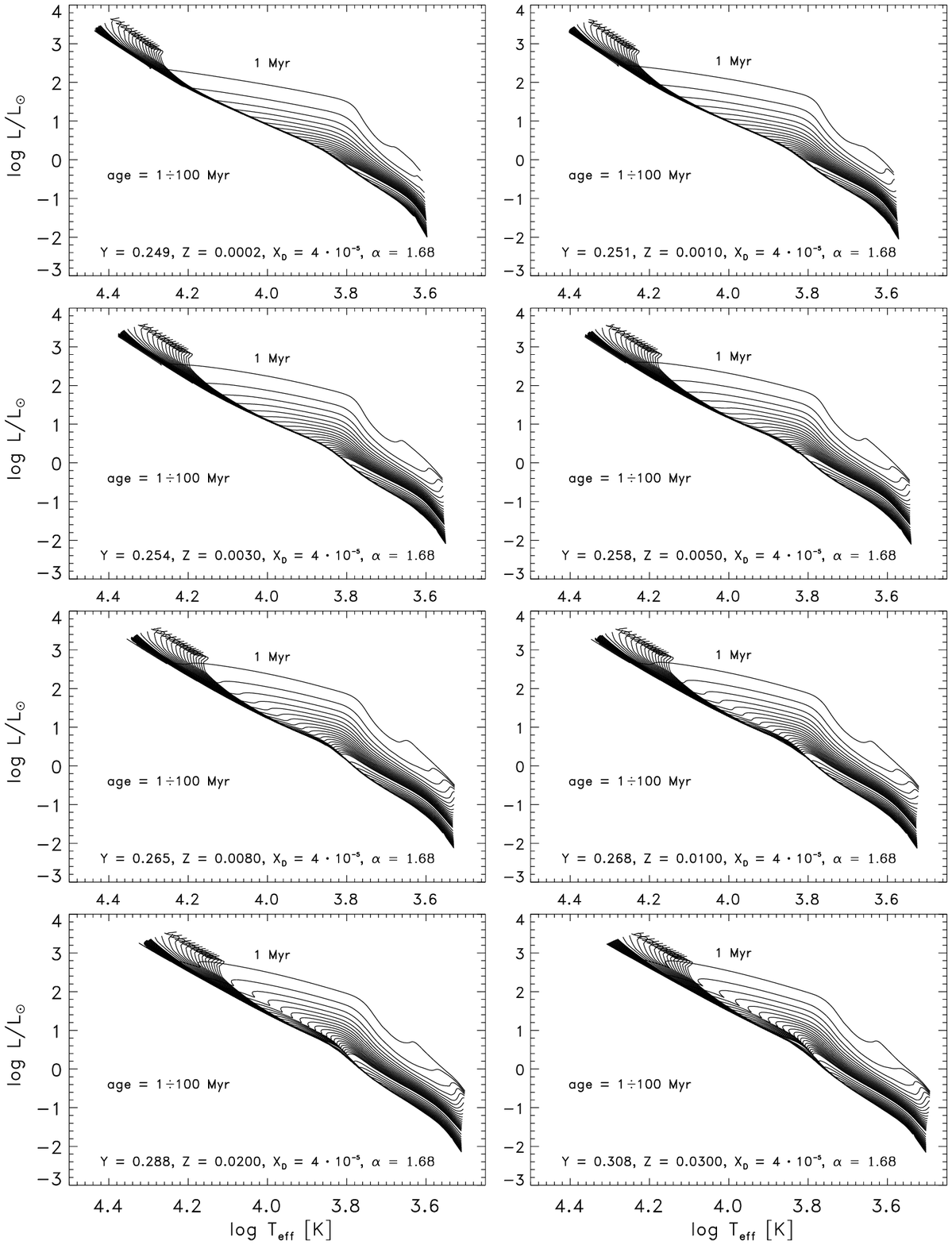}
\caption{Theoretical isochrones in the range $1-100$ Myr for the labelled chemical compositions with $\alpha=1.68$ and X$_{\mathrm{D}}=4\cdot 10^{-5}$.}
\label{fig:multi_iso}
\end{figure*}

Isochrones for ages from 1 Myr to 100~Myr are available; the age spacing is 1 Myr for ages $1-20$ Myr and 5 Myr between 20~Myr and 100 Myr.

Figures \ref{fig:multi_track} and \ref{fig:multi_iso} show a few examples of complete sets of evolutionary tracks and isochrones, respectively, as extracted from our database in the ($\log \mathrm{T}_{\mathrm{eff}}$, $\log \mathrm{L/L}_\odot$) theoretical plane. Figure \ref{fig:multi_track} also shows the isochrones of 1 Myr, which, as discussed in Sect. \ref{sec:initial}, represent a first approximation of the birthline; the dashed part of the tracks correspond to ages younger than 1 Myr, those more affected by theoretical uncertainties.

Models and isochrones are already available in the theoretical plane ($\log \mathrm{T}_{\mathrm{eff}}$, $\log \mathrm{L/L}_\odot$) and will be available in several photometric systems in the next future. The files of each track list: model number, $\log \mathrm{t\,(yr)}$, central abundance by mass of hydrogen, $\log \mathrm{L/L}_\odot$, $\log \mathrm{T}_{\mathrm{eff}}$, central temperature and density ($\log\mathrm{T}_{\mathrm{c}}$ and $\log \rho_{\mathrm{c}}$, respectively), mass of the convective core M$_{\mathrm{cc}}$, contribution to the total luminosity of the proton-proton chain ($\mathrm{L}_{pp}$/$\mathrm{L}_{\mathrm{tot}}$) and \textit{CNO} cycle ($\mathrm{L}_{CNO}$/$\mathrm{L}_{\mathrm{tot}}$) burning and of the gravitational energy ($\mathrm{L}_{\mathrm{g}}$/$\mathrm{L}_{\mathrm{tot}}$), while the files of the isochrone list $\log \mathrm{L/L}_\odot$, $\log \mathrm{T}_{\mathrm{eff}}$, and the mass of the model (M/M$_\odot$).
  
\section{Conclusions}
A growing amount of data of young stellar systems has prompted renewed interest in modelling the initial phase of stellar evolution.

The present paper describes a new set of pre-MS tracks and isochrones, which relies on state-of-the-art input physics (EOS, radiative and conductive opacity, atmospheric models and nuclear cross-section). To provide the astronomical community with a useful and versatile theoretical tool for the interpretation of observational data, the models have been computed for a very large and fine grid of chemical compositions. We have evaluated model data for 19 metallicities, ranging from $\mathrm{Z}=0.0002$ to 0.03, and three different helium abundances for each Z. For a fixed chemical composition, we have made available 43 evolutionary tracks computed with three different values of the mixing-length parameter $\alpha=1.2, 1.68$, and 1.9, in the mass range $0.2-7$~M$_\odot$, and 36 isochrones, in the range $1-100$ Myr. For $\mathrm{Z}\ge 0.008$, two different initial abundances of deuterium have been adopted. The database is available on the web\footnote{ \url{http://astro.df.unipi.it/stellar-models/}}. Models are compared in detail with other computations available in the literature, while the comparison with selected observational data shows good agreement.

\begin{acknowledgements}
We are grateful to Ines Brott who kindly sent us the PHOENIX atmospheric structures we used to build the boundary conditions, and to Marcella Di Criscienzo who sent us the recent ATON pre-MS tracks. It is a great pleasure to thank Emiliano Gregori and Matteo Dell'Omodarme for their invaluable help in computer programming, Mario Gennaro and Giada Valle for their contribution to the updating of the code, and Wolfgang Brandner, Michele Cignoni, Nicola Da Rio, Dimitrios Gouliermis, and Boike Rochau for useful and pleasant discussions on pre-MS evolution. This work has been supported by PRIN-MIUR 2007 ({\em Multiple stellar populations in globular clusters: census, characterizations and origin}, PI G. Piotto) 
\end{acknowledgements}

\bibliographystyle{aa} 
\bibliography{bibliografia_tutti}

\end{document}